\documentclass[a4paper, 11pt]{article}

\renewcommand{\thefootnote}{\fnsymbol{footnote}}

\usepackage{amsbsy,amssymb,latexsym,amsfonts,amsmath}
\usepackage{dsfont,amssymb,booktabs}
\usepackage{mathrsfs}
\usepackage{cite}
\usepackage{bm}
\usepackage{here}
\usepackage{amsmath}
\usepackage{cases}
\usepackage {empheq}
\usepackage{color}
\usepackage[dvipdfmx]{graphicx}
\usepackage{braket}
\usepackage{slashed}
\usepackage{cancel}
\usepackage{longtable}

\numberwithin{equation}{section}
\makeatletter
\def\EqNumText{\refstepcounter{equation}\cdots\tagform@\theequation}%
\makeatother

\newcommand{\bel}[1]{\begin{equation}\label{#1}}                     
\newcommand{\bal}[1]{\begin{eqnarray}\label{#1}}                     
\newcommand{\be}{\begin{equation}}
\newcommand{\ee}{\end{equation}}

\newcommand{\R}{\boldsymbol{\mathrm{R}}}
\newcommand{\T}{\boldsymbol{\mathrm{T}}}

\newcommand\BbbGamma{\reflectbox{\rotatebox[origin=c]{180}{$\mathds L$}}}
\newcommand\BbbGammaVar{\reflectbox{\rotatebox[origin=c]{180}{$\mathbb L$}}}

\title{ 
	\vspace{-15mm}
	\begin{flushright}
		\normalsize
		~~~~
		KEK-TH-2503\\
	\end{flushright}
	\vspace{10mm}
	\textbf{New non-supersymmetric heterotic string theory with reduced rank and exponential suppression of the cosmological constant 
	}\vspace{5mm}
} 

\author{Sota Nakajima\thanks{e-mail: snakajim@post.kek.jp}\vspace{5mm}\\
\normalsize{\it Theory Center, IPNS, High Energy Accelerator Research Organization (KEK),}\\
\normalsize{\it\quad 1-1 Oho, Tsukuba, Ibaraki 305-0801, Japan}}
\date{} 

\begin{document}
	\maketitle
	
	\begin{abstract}
		We study the heterotic asymmetric orbifold model in which supersymmetry is broken by the stringy Schark-Schwarz mechanism. This model is a natural non-supersymmetric extension of CHL strings and can also be interpreted as the interpolating model between the $E_{8}\times E'_{8}$ theory and the non-supersymmetric $E_{8}$ theory. The enhancement of gauge groups, of which the rank is reduced to $8+d$, is explored. In particular, the enhancement to non-simply-laced groups is possible with $d\geq 2$, as well as in the CHL model. We also give the conditions that the massless matter spectrum must satisfy. Moreover, the one-loop cosmological constant is evaluated in the regime where supersymmetry is asymptotically restored, and we show that the exponential suppression can occur unless $d=1$.
	\end{abstract}
\thispagestyle{empty}
\clearpage
\addtocounter{page}{-1}
\newpage
\tableofcontents
\clearpage
\renewcommand{\thefootnote}{\arabic{footnote}}
\setcounter{footnote}{0}
\section{Introduction}\label{sec: introduction}
The landscape of non-supersymmetric (heterotic) string vacua is larger than that of supersymmetric ones. In ten dimensions, for instance, there are only two supersymmetric heterotic string theories: the $E_{8}\times E'_{8}$ theory and the $SO(32)$ theory\cite{Gross:1984dd}. That follows from the fact that only two types of Euclidean even self-dual lattices exist in $16$ dimensions, which are the $E_{8}\times E'_{8}$ lattice and the $Spin(32)/\mathbb{Z}_2$ lattice. On the other hand, by starting from supersymmetric theories and $\mathbb{Z}_{2}$-orbifolding by freely acting shifts on $T_{L}^{16}$ associated with a $2\pi$ spatial rotation, various ten-dimensional heterotic vacua without spacetime supersymmetry can be realized\cite{Dixon:1986iz,AlvarezGaume:1986jb}\footnote{This construction can be generalized to lower dimensional cases just by extending the shifts acting on $T_{L}^{16}$ to on $T_{L}^{16}\times T^d$\cite{Ginsparg:1986wr}.}. In the language of the fermionic formulation, this kind of construction corresponds to introducing a discrete torsion\cite{Vafa:1986wx}.
Those non-supersymmetric theories are classified into six types depending on the independent choices of shift vectors of order $2$: the $SO(16)\times SO(16)$ model, the $SO(32)$ model, the $SO(8)\times SO(24)$ model, the $U(16)$ model, the $SO(16)\times E_8$ model and the $\left( E_{7}\times SU(2) \right)^2$ model. Among them, only the $SO(16)\times SO(16)$ model is tachyon-free. That is a reason why most studies about non-supersymmetric heterotic string phenomenology focus on the $SO(16)\times SO(16)$ model (e.g. in refs. \cite{Blaszczyk:2014qoa,Hamada:2015ria,Perez-Martinez:2021zjj,Baykara:2022cwj,Ashfaque:2015vta}), although it was also proposed that phenomenologically viable models can be obtained by starting from ten-dimensional tachyonic string vacua\cite{Faraggi:2019drl,Faraggi:2019fap,Faraggi:2020hpy,Faraggi:2020wej,Faraggi:2020wld,Faraggi:2021mws}.
By the construction, the $10$D non-supersymmetric theories are discretely related to the 10D supersymmetric ones. 
They can also be connected smoothly via lower dimensional non-supersymmetric models so-called interpolating models \cite{Itoyama:1986ei,Blum:1997gw} which are constructed by the stringy Scherk-Schwarz compactification\cite{Scherk:1978ta,Rohm:1983aq,Kounnas:1989dk}. The supersymmetry breaking scale $m_{3/2}$ is controlled by a radius $R$ of the direction accompanied by the Scherk-Schwarz mechanism as $m_{3/2}\sim 1/R$. Thus, the supersymmetry is restored as $R$ grows, while a non-supersymmetric string model in higher dimensions is reproduced in the limit $R\to 0$. In that sense, these kinds of models interpolate between supersymmetric and non-supersymmetric string vacua. Such models are preferred when a top-down scenario with high-energy supersymmetry breaking is considered (e.g. see \cite{Faraggi:2009xy,Florakis:2021bws,Florakis:2022avh,Abel:2015oxa,Aaronson:2016kjm,Abel:2017vos}). A reason for that is that in the regime with supersymmetry restored, the asymptotic behavior of one-loop cosmological constant takes the form\cite{Itoyama:1986ei,Itoyama:1987rc}\footnote{Through this paper, we set $\alpha'=1$.}
\begin{align}
\label{eq: cc nf-nb}
	\Lambda^{(10-d)}\sim \frac{\xi }{R^{10-d}}\left( n_{F}-n_{B} \right) +\mathcal{O}\left( e^{-R}\right),
\end{align}
where $\xi>0$ is a constant and $n_{F}$ ($n_{B}$) represents the degrees of freedom of massless fermions (bosons). The formula \eqref{eq: cc nf-nb} indicates that the cosmological constant is exponentially suppressed if $n_{F}-n_{B}=0$. Namely, interpolating models with Bose-Fermi degeneracy at the massless level can avoid the instability arising from one-loop dilaton tadpoles. In fact, there are some points in the moduli space where $n_{F}-n_{B}=0$ is realized not only in closed string theories\cite{Kounnas:2016gmz,Kounnas:2017mad,Coudarchet:2017pie,Coudarchet:2018ztz,Itoyama:2019yst,Itoyama:2020ifw,Itoyama:2021fwc,Itoyama:2021itj,Itoyama:2021kxp,Koga:2022qch,Florakis:2016ani,Abel:2017rch} but also in open string theories\cite{Abel:2020ldo,Coudarchet:2020sjw}. 
Note that whether this suppression is achieved or not depends on the spectrum of only massless states. It is known that it is possible to construct type II asymmetric orbifold models without supersymmetry but with Bose-Fermi degeneracy at each mass level, and thus the cosmological constants can strictly vanish at one-loop even in non-supersymmetric cases\cite{Kachru:1998hd,Shiu:1998he,Satoh:2015nlc,Sugawara:2016lpa}.

All the non-supersymmetric models mentioned above are constructed by $\mathbb{Z}_{2}$ freely acting orbifolds, and the rank of (left-moving) gauge groups is always $16+d$ with compact dimension $d$.
By using the outer automorphism of the $E_{8}\times E'_{8}$ lattice (denoted by $\Gamma_{E_{8}}\oplus \Gamma_{E'_{8}}$) instead of freely acting shifts, it is possible to present heterotic strings in which the rank of gauge groups is reduced to $8+d$. One has to note that orbifolding by only this automorphism, which we denote by $\R$ in this paper, gives the original $E_{8}\times E'_{8}$ theory since the twisted sector reproduces the spectrum projected out by $\R$\footnote{This fact is valid not only for $\R$ but also for a $2\pi$ spatial rotation. Twisting one of the two supersymmetric heterotic theories only by a freely acting shift can give the other heterotic theory \cite{Dixon:1986jc}.}. In order to obtain a different model, hence, we need to define a $\mathbb{Z}_{2}$-orbifold twist by combining $\R$ with a nontrivial $\mathbb{Z}_{2}$ generator.
For instance, using a $2\pi$ spatial rotation as an additional $\mathbb{Z}_2$ generator, we obtain the reduced rank non-supersymmetric $E_{8}$ model in ten dimensions \cite{Kawai:1986vd}, which we review in appendix \ref{app: non-susy E8}. In lower dimensions, one can use a half-shift along one of the compact directions as an additional $\mathbb{Z}_{2}$ generator and build a reduced rank model with all supercharges preserved which is called the CHL string model \cite{Chaudhuri:1995fk}, which is originally built by using the fermionic construction \cite{Kawai:1986ah,Kawai:1986vd}. One of the interesting features of CHL strings is that gauge groups can be enhanced to non-simply-laced ones if the compact dimension $d$ is greater than or equal to $2$. Recently, in ref. \cite{Font:2021uyw}, the algorithms to explore the moduli space of the CHL strings have been developed and the maximal enhancements of gauge groups have been completely identified for $d=1,2$. 
In this paper, we investigate the non-supersymmetric reduced rank models, which we call the non-supersymmetric CHL models, constructed by the asymmetric orbifold with the $\mathbb{Z}_2$-twist given by the product of the three $\mathbb{Z}_{2}$ generators: the outer automorphism $\R$ of $\Gamma_{E'_{8}}\times \Gamma_{E'_{8}}$, the half-shift $\T$ along a compatified direction and a $2\pi$ spatial rotation $(-1)^{F}$\footnote{In ref. \cite{Partouche:2020swy}, another way to build reduced rank heterotic orbifold models has been proposed by using the $SO(2n+1)$ characters.}. We will show that in the reduced rank model, the formula \eqref{eq: cc nf-nb} is still valid, and the cosmological constant can be exponentially suppressed.

As mentioned above, the ingredients that we need to construct the orbifold models in this paper are only the following three types of the $\mathbb{Z}_2$ generators:
\begin{itemize}
	\item outer automorphism of the $E_{8}\times E'_{8}$ lattice, i.e., interchange of the two $E_{8}$ factors:
	\begin{align}
		\R:~\Gamma_{E_{8}}\oplus\Gamma_{E'_{8}} \to \Gamma_{E'_{8}}\oplus\Gamma_{E_{8}}.
	\end{align}
	\item half translation along a compact direction $x^1$ with a periodicity $x^1\sim x^1+2\pi R$:
	\begin{align}
	\T:~x^{1}\to x^{1}+\pi R.
	\end{align}
	\item $2\pi$ spatial rotation: $(-1)^F$ where $F$ is the spacetime fermion number.
\end{itemize}
We here summarize reduced rank heterotic string models depending on the combinations of the above $\mathbb{Z}_{2}$ generators:
\begin{itemize}
	\item $\R\T$: the CHL string (see section \ref{sec: CHL string}).
	\item $\R(-1)^{F}$: the non-supersymmetric $E_{8}$ string (see appendix \ref{app: non-susy E8}).
	\item $\R\T(-1)^{F}$: the non-supersymmetric CHL string (see section \ref{sec: non-susy CHL string}).
\end{itemize}
This paper is organized as follows. In section \ref{sec: CHL string}, we review the CHL model, including the partition function, the momentum lattice and the massless spectrum. Those are useful when we study the non-supersymmetric CHL model. We will see why the enhancement to non-simply-laced groups is possible in the reduced rank model.
In section \ref{sec: non-susy CHL string}, we move to consider the non-supersymmetric CHL model. The partition function can be expressed in terms of the building blocks of the CHL string. We show that this non-supersymmetric model interpolates between the $E_{8}\times E'_{8}$ theory and the $E_{8}$ theory. Comparing in the CHL model, moreover, we identify the conditions for $U(1)_{L}^{16}\times U(1)_{l}^{d}$ to be enhanced a non-Abelian group and the formulae that matter contents in the massless spectrum must satisfy. In particular, some explicit examples of symmetry enhancement to non-simply-laced algebra are given in the case $d=2$.
We then evaluate the one-loop cosmological constant of the non-supersymmetric CHL string in the regime $R\gg1$ and show that the exponential suppression may occur for $d\geq 2$ in section \ref{sec: cosmological const}. Indeed, we give examples of the moduli that cause the suppression in the cases $d=2$ and $d=3$.
The summary and discussion are given in section \ref{sec: summary}. 

\section{Reveiw of CHL strings}\label{sec: CHL string}
In this section, we review the CHL string model, i.e., an asymmetric orbifold model with $\R\T$. We will give the conditions on the internal momenta of massless states and see that the symmetry enhancement to non-simply-laced algebras is possible if $d\geq 2$ (see \cite{Font:2021uyw} for more detail).

\subsection{Partition function and momentum lattice}\label{subsec: partition function CHL string}
The starting point is the $E_8\times E'_8$ heterotic model compactified on $T^d$, in which we have $(16+d)\times d$ moduli: the metric $G_{mn}$, the anti-symmetric tensor $B_{mn}$ and the Wilson lines $A_{m}^{A}$, where the indices $m,n=1,\cdots,d$ and $A=1,\cdots,16$ represent the directions along $T^d$ and $T_{L}^{16}=\mathbb{R}^{16}/( \Gamma_{E_{8}}\oplus \Gamma_{E'_{8}})$ respectively.
The internal momenta, which are identified as elements of even self-dual lattice with the signature $(16+d,d)$ so-called Narain lattice (denoted by $\BbbGammaVar$), can be written as \cite{Narain:1985jj,Narain:1986am}
\begin{subequations}
\begin{align}
\ell_{L}^{A}&=\Pi^{A}-w^{m}A_{m}^{A},\\
p_{L,a}&=\frac{1}{\sqrt{2}}\left(  \Pi^{A}A_{m}^{A}+w^{n}\left( G-C\right)_{nm}+n_{m}  \right)  e^{*m} _{a}, \\
p_{R,a}&=\frac{1}{\sqrt{2}}\left(  \Pi^{A}A_{m}^{A}-w^{n}\left( G+C\right)_{nm}+n_{m}  \right)  e^{*m} _{a},
\end{align}
\end{subequations}
where $\Pi^{A}=\left( \pi^I,\pi'^{I}\right) \in\Gamma_{E_{8}}\oplus \Gamma_{E'_{8}}$ with $I=1,\cdots,8$. Here $C_{mn}:=-B_{mn}+\frac{1}{2}A^{A}_{m}A^{A}_{n}$ and $e^{a}_{m}$ is a vierbein on $T^d$ satisfying
\begin{align}
e^{a}_{m}e^{a}_{n}=G_{mn},~~~e^{a}_{m}e^{*n}_{a}=\delta_{m}^{n},~~~e^{*m}_{a}e^{*n}_{a}=(G^{-1})^{mn}.
\end{align}
In order to be consistent with the orbifold symmetry $\R$, the components of the Wilson lines must satisfy
\begin{align}
\label{eq: consistent WLs}
A_{m}^{I}=A_{m}^{I+8}=a_{m}^{I}\in M_{d\times 16}(\mathbb{R}),
\end{align}
where $M_{d\times d'}(\mathbb{R})$ represents a set of $d\times d'$ matrices with real components.
Since $\pi^{I}$ and $\pi'^{I}$ are interchanged under $\R$, it is convenient to change the basis as
\begin{align}
\ell_{\pm}^{I}=\frac{1}{\sqrt{2}}\left( \ell_{L}^{I}\pm\ell_{L}^{I+8}\right).
\end{align}
With the background \eqref{eq: consistent WLs}, the internal momenta are then
\begin{subequations}
	\label{eq: momenta changed}
	\begin{align}
	\label{eq: ell-}
	\ell_{-}&=\frac{1}{\sqrt{2}}\pi_{-},\\
	\label{eq: ell+}
	\ell_{+}&=\frac{1}{\sqrt{2}}\left( \pi_{+}-2wa\right) ,\\
	\label{eq: pL}
	p_{L}&=\frac{1}{\sqrt{2}}\left(\pi_{+} \cdot a+w\left( 2G-E\right)+n  \right)  e^{-t},\\
	\label{eq: pR}
	p_{R}&=\frac{1}{\sqrt{2}}\left(  \pi_{+} \cdot a-wE+n \right)  e^{-t}.
	\end{align}
\end{subequations}
In the above expression, we adopt the notations of the row vectors
\begin{align}
w\in\mathbb{Z}^{d},~~~n\in\mathbb{Z}^{d},~~~\pi_{\pm}:=\pi\pm \pi'\in\Gamma_{E_{8}},
\end{align}
and the matrices\footnote{To be precise, the matrix $G$ is more restricted since it must be positive definite.}
\begin{align} 
a\in M_{d\times 16}(\mathbb{R}),~~~G=G^t\in M_{d\times d}(\mathbb{R}),~~~E:=G-B+aa^t\in M_{d\times d}(\mathbb{R}).
\end{align}
Note that only the component $\ell_{-}$ is reflected under $\R$:
\begin{align}
\R:~\left(\ell_{-},\ell_{+},p_{L};p_{R} \right) \to \left(-\ell_{-},\ell_{+},p_{L};p_{R} \right).
\end{align}
By introducing the generalized vierbein, we can express the momenta \eqref{eq: momenta changed} more simply:
\begin{align}
\left(\ell_{-}, \ell_{+}, p_L;p_{R} \right) = \left( \pi_{-},\pi_{+}, w,n\right) \mathcal{E}\left(e,B,A \right),
\end{align}
where the generalized vierbein $\mathcal{E}$ is a $(16+2d)\times (16+2d)$ matrix
\begin{align}
\mathcal{E}\left(e,B,a \right) =
\left(\begin{array}{cc}
\frac{1}{\sqrt{2}}\boldsymbol{1}_{8}& 0  \\
0&   \mathcal{E}^{(d)}\left(e,B,a \right)
\end{array}
\right),
\end{align}
with the $(8+2d)\times (8+2d)$ matrix $\mathcal{E}^{(d)}$ given by
\begin{align}
\mathcal{E}^{(d)}\left(e,B,a \right) =\frac{1}{\sqrt{2}} 
\left(\begin{array}{ccc}
\boldsymbol{1}_{8}& a^t e^{-t}& a^t e^{-t} \\
-2a& \left( G-C\right)e^{-t} & -\left( G+C\right)e^{-t}\\
0& e^{-t}&e^{-t}
\end{array}
\right).
\end{align}
It is convenient to define a $(8+2d)$-dimensional lattice $\Gamma^{8+d,d}$ as
\begin{align}
	\Gamma^{8+d,d}:=\left\lbrace P=\left( \pi_{+}, w, n\right) \mathcal{E}^{(d)}\left(e,B,a \right) \left|~ \pi_{+}\in \Gamma_{E_{8}},~w,n\in\mathbb{Z} ^{d}\right.  \right\rbrace.
\end{align}
The Lorentzian inner product in $\Gamma^{8+d,d}$ is computed by the Narain-like metric 
\begin{align}
\label{eq: Narain metric}
\mathcal{E}^{(d)}\left(e,B,a \right)\eta^{(8+d,d)}\mathcal{E}^{(d)t}\left(e,B,a \right)=\left(\begin{array}{ccc}
\frac{1}{2}\boldsymbol{1}_{8}& 0& 0 \\
0&0 & \boldsymbol{1}_{d}\\
0& \boldsymbol{1}_{d}&0
\end{array}
\right),
\end{align}
where $\eta^{(8+d,d)}=\mathrm{diag}(\boldsymbol{1}_{8+d},-\boldsymbol{1}_{d})$. Note that $\Gamma^{8+d,d}$ is not even lattice unlike $\BbbGamma$:
\begin{align}
\label{eq: norm}
	\ell_{+}^2+p_{L}^2-p_{R}^2=\frac{1}{2}\pi_{+}^{2}+2wn^t\in\mathbb{Z}.
\end{align}

The partition function of the CHL model is expressed as 
\begin{align}
Z_{\text{CHL}}^{(10-d)}=\frac{1}{2}Z_{B}^{(8-d)}\left( \bar{V}_{8}-\bar{S}_{8} \right) \sum_{a,b=0,1} Z(g^a,g^b),
\end{align}
where $Z_{B}^{(8-d)}:=\tau_{2}^{-(8-d)/2}(\eta\bar{\eta})^{-(8-d)}$ is the contribution from the world-sheet bosonic fields propagating in the spacetime dimensions, and $\bar{V}_{8}$ and $\bar{S}_{8}$ are the conjugacy classes of the $SO(8)$ vector and spinor (see appendix \ref{app: lattice characters}) which are the contribution from the right-moving fermionic fields.
Each of the building blocks $Z(g^a,g^b)$ can be identified as (see appendix in ref. \cite{Font:2021uyw} for detail)
\begin{subequations}
	\label{eq: conformal blocks CHL}
\begin{align}
Z(1,1)&=\frac{1}{\bar{\eta}^{d}\eta^{16+d}}\sum_{\left(\ell_{-},P \right)\in \BbbGamma}q^{\frac{1}{2}\left( \ell_{-}^2+P_{L}^2\right) }\bar{q}^{\frac{1}{2}p_{R}^2},\\
Z(1,g)&=\frac{1}{\bar{\eta}^{d}\eta^{16+d}}\left( \frac{2\eta^3}{\vartheta_{2}}\right)^{4} \sum_{P\in \mathbb{I}}q^{\frac{1}{2}P_{L}^2 }\bar{q}^{\frac{1}{2}p_{R}^2}e^{2\pi i P\cdot V},\\
Z(g,1)&=\frac{1}{\bar{\eta}^{d}\eta^{16+d}}\left( \frac{\eta^3}{\vartheta_{4}}\right)^{4} \sum_{P\in \mathbb{I}^{*} }q^{\frac{1}{2}\left( P_{L}+V_{L}\right) ^2 }\bar{q}^{\frac{1}{2}\left( p_{R}+v_{R}\right)^2},\\
Z(g,g)&=\frac{1}{\bar{\eta}^{d}\eta^{16+d}}\left( \frac{e^{\frac{\pi i}{4}}\eta^3}{\vartheta_{3}}\right)^{4} \sum_{P\in \mathbb{I}^{*} }q^{\frac{1}{2}\left( P_{L}+V_{L}\right) ^2 }\bar{q}^{\frac{1}{2}\left( p_{R}+v_{R}\right) ^2}e^{\pi i \left( P+V\right)^2 }.
\end{align}
\end{subequations}
The first and second arguments of $Z(g^a,g^b)$ indicate the twists along the spatial and time directions on the world-sheet respectively, and $Z(g,1)+Z(g,g)$ is the contribution from the twisted sector which is necessary for $Z_{\text{CHL}}^{(d)}$ to be modular invariant\cite{Dixon:1985jw,Dixon:1986jc}.
Here $\mathbb{I}$ and $\mathbb{I}^{*}$ are the invariant lattice under $\R$ and its dual, which are identified below, and $V=\left(V_{L},v_R \right)$ is a shift vector of order $2$ on $\Gamma^{8+d,d}$, which can be expressed as
\begin{align}
	V=\frac{1}{2}\left( \hat{\pi}_{+},\hat{w},\hat{n}\right)\mathcal{E}^{(d)}\left( e,B,A\right),~~~\text{for }~\hat{\pi}_{+}\in\Gamma_{E_{8}},~\hat{w},\hat{n}\in\mathbb{Z}^{d}.
\end{align}
In order for the total partition function to be modular invariant\footnote{To be precise, $Z(g,1)$ must be invariant under $\tau\to\tau+2$.}, the shift vector $V$ must be chosen such that 
\begin{align}
\label{eq: constraint on V}
	\left( P+V\right)^2\in \mathbb{Z},~~~\text{for any $P\in\mathbb{I}^{*}$}.
\end{align}
In this paper, we choose $\hat{\pi}_{+}=0$, $\hat{n}=0$ and $\hat{w}=(1,0^{d-1})$ so that $V$ generates the half-shift along $x^1$.

The invariant lattice $\mathbb{I}$ is defined by a subset of $\BbbGamma$ in which elements are invariant under the exchange $\pi\leftrightarrow\pi'$, i.e., $\pi_{-}=0$ and $\pi_{+}\in 2\Gamma_{E_{8}}$. Thus, we can express $\mathbb{I}$ in terms of the vierbein $\mathcal{E}^{(d)}$:
\begin{align}
\mathbb{I}=\left\lbrace P=\left( \pi_{+}, w, n\right)\mathcal{E}^{(d)}\left(e,B,a \right) \left|~ \pi_{+}\in 2\Gamma_{E_{8}},~w,n\in\mathbb{Z} ^{d}\right.  \right\rbrace.
\end{align}
Since the inner product in $\mathbb{I}$ is defined by the metric \eqref{eq: Narain metric}, we can identify the dual lattice $\mathbb{I}^{*}$ as
\begin{align}
\mathbb{I}^{*}=\Gamma^{8+d,d}=\left\lbrace P=\left( \pi_{+}, w, n\right)\mathcal{E}^{(d)}\left(e,B,a \right) \left|~ \pi_{+}\in \Gamma_{E_{8}},~w,n\in\mathbb{Z} ^{d}\right.  \right\rbrace.
\end{align}
The internal momenta of the twisted states are shifted by $V$, and the momentum lattice in the twisted sector is given by
\begin{align}
\mathbb{I}^{*}+V=\left\lbrace P=\left( \pi_{+}, w, n\right) \mathcal{E}^{(d)}\left(e,B,a \right) \left|~ \pi_{+}\in \Gamma_{E_{8}},~w\in\mathbb{Z}^{d}+\frac{\hat{w}}{2},~n\in\mathbb{Z}^{d}\right.  \right\rbrace.
\end{align}
A more detailed study of the momentum lattice of CHL strings was done in ref. \cite{Mikhailov:1998si}.

\subsection{Massless spectrum}\label{subsec: massless spectrum CHL string}
Let us study the spectrum of the CHL string.
As well as for the internal momenta, it is convenient to define
\begin{align}
	X_{\pm}^{I}:=\frac{1}{\sqrt{2}}\left(X_{L}^{I}\pm X_{L}^{I+8} \right).
\end{align}
The transformation law for the oscillators under $\R$ is the same as for the momenta $(\ell_{-},\ell_{+},p_{L};p_{R})$:
\begin{align}
\label{eq: R image oscillators}
	\R:~\left(\alpha_{-}^{I},\alpha_{+}^{I},\alpha^{\mu} \right) \to \left(-\alpha_{-}^{I},\alpha_{+}^{I},\alpha^{\mu} \right),
\end{align}
where $\alpha_{\pm}^{I}$ denotes the oscillators for $X_{\pm}^{I}$ and $\alpha^{\mu}$ for $X^{\mu}$ with $\mu=2,\cdots,9$ being the indices of the spacetime transverse directions.
The spectrum in the untwisted sector, which is invariant under $\R\T$, consists of the following combinations of states:
\begin{align}
\ket{\varphi}_{\text{untw}}=\frac{1}{\sqrt{2}}\left( \alpha\ket{w,n,\pi,\pi'}+(-)^{n_1}g(\alpha)\ket{w,n,\pi',\pi}\right) ,
\end{align}
where $\alpha$ denotes any possible combinations of the oscillators and $g(\alpha)$ its image under $\R$, given by \eqref{eq: R image oscillators}.
Note that the untwisted states without oscillator excitation and with $\pi=\pi'$ must have $n_1$ even. As we will see later, this fact is the reason why the gauge symmetry can not be enhanced to non-ADE algebras in the case $d=1$.

In the twisted sector, $X_{-}^{I}(\sigma+t)$ is anti-periodic under $\sigma\to\sigma+2\pi$, up to elements of $\Gamma_{E_{8}}$ multiplied by $1/\sqrt{2}$. So, the oscillator excitation for the $X_{-}^{I}$-directions is given by $\alpha^{I}_{-,r}$ with $r\in\mathbb{Z}+1/2$.
The orbifold projection in the twisted sector can be figured out by the partition function:
\begin{align}
\frac{1}{2}\left( Z(g,1)+Z(g,g)\right) &=\frac{1}{\bar{\eta}^d\eta^{16+d}}\sum_{P\in \mathbb{I}^{*}+V }q^{\frac{1}{2}P_{L}^2 }\bar{q}^{\frac{1}{2}p_{R}^2}
\frac{1}{2}\left( \left( \frac{\eta^3}{\vartheta_{4}}\right)^{4}- \left( \frac{\eta^3}{\vartheta_{3}}\right)^{4} e^{\pi i \left( \frac{\pi_{+}^2}{2}+n_{1}\right) }\right)
\nonumber\\
&=\frac{1}{\bar{\eta}^d\eta^{16+d}}\sum_{s=0,1}
 \sum_{\substack{P\in \mathbb{I}^{*}+V \\\frac{\pi_{+}^2}{2}+n_{1}\in 2\mathbb{Z}+s}}q^{\frac{1}{2}P_{L}^2 }\bar{q}^{\frac{1}{2}p_{R}^2}
F\left( q,s\right),
\end{align}
where
\begin{align}
	F\left( q,s\right):=\frac{1}{2}\left( \left( \frac{\eta^3}{\vartheta_{4}}\right)^{4}-e^{\pi i s} \left( \frac{\eta^3}{\vartheta_{3}}\right)^{4} \right).
\end{align}
Note that $s=0,1$ specifies whether $P^2$ is even or odd, i.e.,
\begin{align}
\label{eq: definition of s}
	P^2=P_{L}^2-p_{R}^2\stackrel{2}{=}\frac{\pi_{+}^2}{2}+n_{1}\stackrel{2}{=}s,
\end{align}
where the symbol ``$\stackrel{n}{=}$" indicates that the equality holds for mod $n$.
The $q$-expansion of $F(q,s)/\eta^8$, which gives the contribution from the oscillators $\alpha_{-}^{I}$ with the orbifold projection, is
\begin{align}
\frac{F(q,s)}{\eta^8}=\frac{1}{2}\left( 
\left( \prod_{n=1}^{\infty}\sum_{N_{n}=0}^{\infty} \left( q^{n-\frac{1}{2}}\right)^{N_{n}}\right) ^8
-e^{\pi i s}\left( \prod_{n=1}^{\infty}\sum_{N_{n}=0}^{\infty} \left( -q^{n-\frac{1}{2}}\right)^{N_{n}}\right) ^8\right) ,
\end{align}
and thus we find that the twisted states take the form
\begin{align}
\label{eq: twisted states CHL}
\ket{\varphi}_{\text{tw}}=\frac{1}{\sqrt{2}}\left( \alpha\ket{\pi_{+},w,n}+(-1)^{s+1}g(\alpha)\ket{\pi_{+},w,n}\right) ,
\end{align}
where $g(\alpha)$ is given by \eqref{eq: R image oscillators}. Let $N_{-}$ be the total number of $\alpha_{-,r}^{I}$ acting on $\ket{\pi_{+},w,n}$. The projection \eqref{eq: twisted states CHL} requires that $N_{-}$ must be even (odd) for $P^2$ odd (even), and as we will see below, indeed, all states satisfying the level-matching condition survive under the projection \eqref{eq: twisted states CHL}.

The mass formula is given by
\begin{subequations}
	\label{eq: mass formula CHL string}
	\begin{align}
	\label{eq: mass formula right}
	M_{R}^{2}&=p_{R}^{2}+2\left( N_{R}+a_{R}\right) ,\\
	\label{eq: mass formula left}
	M_{L}^{2}&=
	\begin{cases}
	P_{L}^{2}+\ell_{-}^2+2\left( N_{L}-1\right)&(\text{untwisted sector})\\
	P_{L}^{2}+2\left( N_{L}-\frac{1}{2}\right)&(\text{twisted sector})\\
	\end{cases} ,
	\end{align}
\end{subequations}
where $a_{R}$ is the zero-point energy for the right-moving states: $a_{R}=-1/2~(0)$ for NS (R) sector. The level-matching condition in the twisted sector reads
\begin{align}
\label{eq: level-matching twisted sector}
	P_{L}^2-p_{R}^2+2\left( N_{L}-\frac{1}{2}-N_{R}-a_{R}\right) =0.
\end{align}
The GSO projection for the right-moving states is performed such that $N_{R}+a_{R}$ is an integer, and hence $P^2$ must be even (odd) for $N_{-}$ odd (even) in order to satisfy the condition \eqref{eq: level-matching twisted sector}. Then, the level-matching condition demands that the twisted states should take the form \eqref{eq: twisted states CHL}.

Taking the GSO projection into account, $M_{R}=0$ is realized only when $N_{R}+a_{R}=0$ and $p_{R}=0$. The latter is explicitly written as
\begin{align}
\label{eq: p_R=0}
	n= -\pi_{+}\cdot a+wE \in\mathbb{Z}^d.
\end{align}
In order for a (left-moving) untwisted state to be massless, the excitation number $N_{L}$ must be either $1$ or $0$;
\begin{itemize}
	\item $N_{L}=1$;\\
	This state is massless if $P_{L}=\ell_{-}=0$, i.e., $\pi=\pi'=w=n=0$. By acting $\alpha_{-1}^{\mu}$ on the ground state, we obtain a gravity multiplet and KK gauge bosons in NS sector and their fermionic partners in R sector. As for the excitation by $\alpha_{-1}^{A}$, only the combinations
	\begin{align}
	\alpha_{-1}^{I}+ \alpha_{-1}^{I+8}
	\end{align}
	survive under the orbifold projection. The other independent eight Cartan states are projected out, and then the rank of gauge groups is reduced to $8+d$.
	
	\item $N_{L}=0$;\\
	With no oscillator excitation, by the mass formula \eqref{eq: mass formula left}, it turns out that $M_{L}^2=0$ requires
	\begin{align}
	\ell_{-}^2+P_{L}^2=2.
	\end{align}
	Since $\ell_{-}^2$ is a non-negative integer, we have the following two types of solutions\footnote{Note that $P_{L}=0$ ($\ell_{-}^2=2$) is excluded since $P_{L}=p_R=0$ implies $\pi_{+}=w=n=0$, but $\ell_{-}^2\in 4\mathbb{Z}$ under $\pi_{+}=0$, which is inconsistent with $\ell_{-}^2=2$.}:
	\begin{align}
	\label{eq: roots untwisted CHL}
		P_{L}^2\stackrel{\eqref{eq: p_R=0}}{=}\frac{1}{2}\pi_{+}^2+2wn=\begin{cases}
		1&\text{with }\left( \pi-\pi'\right)^2=1\\
		2&\text{with }~\pi=\pi'
		\end{cases}.
	\end{align}
	Note that in the second one, $\pi_{+}$ are restricted to $\pi_{+}\in 2\Gamma_{E_{8}}$ and thus $n_1$ must be even. (Recall that states with $N_{L}=0$, $\pi=\pi'$ are projected out if $n_1$ is odd.)
	A set of states with $P_{L}^2=2$ corresponds to long roots of a non-simply-laced algebra. Namely, the enhancement to a non-ADE gauge group is realized if there exist solutions $\left( \pi_{+},w,n\right)$ of $P_{L}^2=2$ satisfying the condition \eqref{eq: p_R=0} for a given set of the moduli.
	One can see that in the case $d=1$, the enhancement to non-simply-laced algebras is impossible since 
	\begin{align}
	\label{eq: long roots d=1}
		\frac{1}{2}\pi_{+}^2+2w^1 n_1\in 4 \mathbb{Z},~~~\text{for }\pi_{+}\in 2\Gamma_{E_{8}},~n_{1}\in 2\mathbb{Z},
	\end{align}
	and $P_{L}^2$ can not be $2$.
	In the case $d\geq 2$, however, in eq. \eqref{eq: roots untwisted CHL} we have $2\sum_{m=2}^{d}w^m n_{m}$ in addition to \eqref{eq: long roots d=1}, and solutions for long roots can be found at appropriate points in the moduli space.
	
\end{itemize}

In the twisted sector, $P_{L}^2$ can not vanish due to $w^1\in\mathbb{Z}+1/2$, and $M_{L}^2=0$ leads to $N_{L}=0$ and $P_{L}^2=1$. Substituting $p_{R}=0$ and $P_{L}^2=1$ into eq. \eqref{eq: norm}, we obtain
\begin{align}
\label{eq: roots twisted CHL}
\frac{1}{2}\pi_{+}^2+2wn=1,~~~\text{for }\pi_{+}\in\Gamma_{E_{8}},~w\in \mathbb{Z}^d+\frac{\hat{w}}{2},~n\in\mathbb{Z}^d.
\end{align}
States with long roots $P_{L}^2=2$ never appear in the massless spectrum of the twisted sector.

Noting $p_{L}=p_{R}+\sqrt{2}we$, one can find an alternative formula of \eqref{eq: roots untwisted CHL} or \eqref{eq: roots twisted CHL}:
\begin{align}
\label{eq: rewritten roots untwisted CHL}
	P_{L}^2=\frac{1}{2}\left( \pi_{+}-2wa\right)^2+2wGw^t=1\text{ or }2.
\end{align}
For long roots, in particular, this formula is expressed as
\begin{align}
	\left( \pi'_{+}-wa\right)^2+wGw^t=1,~~~~\text{for }\pi'_{+}=\frac{\pi_{+}}{2}\in\Gamma_{E_{8}}.
\end{align}
The formula \eqref{eq: rewritten roots untwisted CHL} is more convenient in order to find a finite set of $\left( \pi_{+},w,n\right) $ satisfying $p_{R}=0$ and $P_{L}^2=1,2$.

\section{Non-supersymmetric CHL strings}\label{sec: non-susy CHL string}
In the following sections, we construct the asymmetric orbifold with $\R\T(-1)^F $ and explore the non-supersymmetric CHL model. A point that we have to note is that the $\mathbb{Z}_{2}$-twist acts on bosonic states and fermionic states in different ways. Namely, bosonic states in the untwisted sector are even under $\R\T$ as well in the CHL model, while fermionic states take the combinations such that the eigenvalue under $\R\T$ is $-1$. Thus, the bosonic spectrum in the untwisted sector is the same as in the CHL model, and the fermionic spectrum is provided by the opposite projection.
The twisted sector in this case turns out to contain the scalar and conjugate spinor (co-spinor) conjugacy class of the spacetime $SO(8)$ (see appendix \ref{app: lattice characters}). 

We provide the partition function which is not only useful to explore the spectrum but also necessary to evaluate the cosmological constant in section \ref{sec: cosmological const}. Then we study the massless spectrum of the non-supersymmetric CHL string. The examples for $d=1,2$ are given, and we will explicitly see that the gauge symmetries are enhanced to non-ADE algebras in the case $d=2$.

\subsection{Partition function}
The partition function of the non-supersymmetric CHL string is expressed in terms of the building blocks $Z(g^a,g^b)$ of the CHL string, given by \eqref{eq: conformal blocks CHL}. Since a $2\pi$ spatial rotation $(-1)^{F}$ assigns $\bar{V}_{8}$ and $\bar{S}_{8}$ to $+1$ and $-1$ respectively, the partition function for the untwisted sector takes the form
\begin{align}
	\frac{1}{2}Z_{B}^{(8-d)}\left\lbrace 
	\left( \bar{V}_{8}-\bar{S}_{8}\right)  Z(1,1)+\left( \bar{V}_{8}+\bar{S}_{8}\right)  Z(1,g)\right\rbrace.
\end{align}
From the transformation law \eqref{eq: S-trans characters} of the $SO(2n)$ characters, we see under $\tau \to -1/\tau$,
\begin{align}
	\left( \bar{V}_{8}+\bar{S}_{8}\right)  Z(1,g) \xrightarrow[]{\tau\to-\frac{1}{\tau}} \left( \bar{O}_{8}-\bar{C}_{8}\right)  Z(g,1).
\end{align}
Performing the transformation $\tau \to \tau+1$, we find
\begin{align}
\left( \bar{O}_{8}-\bar{C}_{8}\right)  Z(1,g) \xrightarrow[]{\tau\to\tau+1} -\left( \bar{O}_{8}+\bar{C}_{8}\right)  Z(g,g).
\end{align}
The total partition function is then 
\begin{align}
\label{eq: partition function non-susy CHL}
Z_{\mathcal{N}=0}^{(10-d)}&=\frac{1}{2}Z_{B}^{(8-d)}\left\lbrace 
\left( \bar{V}_{8}-\bar{S}_{8}\right)  Z(1,1)+\left( \bar{V}_{8}+\bar{S}_{8}\right)  Z(1,g)\right.\nonumber \\
&\left. ~~~~~~~~~~~~~~~~
+\left( \bar{O}_{8}-\bar{C}_{8}\right)  Z(g,1)-\left( \bar{O}_{8}+\bar{C}_{8}\right)  Z(g,g)\right\rbrace  \nonumber\\
&=\frac{1}{2}Z_{B}^{(8-d)}\left\lbrace 
\bar{V}_{8}\left( Z(1,1)+Z(1,g)\right) 
-\bar{S}_{8} \left( Z(1,1)-Z(1,g)\right) \right. \nonumber\\
&\left. ~~~~~~~~~~~~~
+\bar{O}_{8}\left( Z(g,1)-Z(g,g)\right) 
-\bar{C}_{8} \left( Z(g,1)+Z(g,g)\right)
\right\rbrace.
\end{align}
One can see that the contribution from $\alpha_{-}^{I}$ in the co-spinor conjugacy class in the twisted sector is $F(q,s)/\eta^8$ as well as in the CHL model, while in the scalar conjugacy class is $F(q,s+1)/\eta^8$.

In order to see that this model interpolates between the $E_{8}\times E'_{8}$ theory and the $E_{8}$ theory, let us restrict our attention to the case $d=1$ without the Wilson line ($a=0$) for simplicity. In the limit $R\to \infty$, the contribution with $w^1\neq 0$ to the partition function is suppressed, which means that the twisted sector vanishes. One can also find that the behavior of the untwisted sector in $R\to\infty$ is given by
\begin{align}
	Z(1,1)&\sim 
	\frac{1}{\bar{\eta}\eta^{17}}\sum_{\Pi\in \Gamma_{E_{8}}\oplus\Gamma_{E'_{8}}} q^{\frac{1}{2}\Pi^2 }\sum_{n\in\mathbb{Z}}e^{-\pi\tau_{2}\left( \frac{n}{R}\right)^2}
	\xrightarrow[]{R\to\infty}
	\frac{R}{\sqrt{\tau_{2}}}\frac{1}{\bar{\eta}\eta^{17}}\sum_{\Pi\in \Gamma_{E_{8}}\oplus\Gamma_{E'_{8}}} q^{\frac{1}{2}\Pi^2 },\\
	Z(1,g)&\sim 
	\frac{1}{\bar{\eta}\eta^{17}}\sum_{\pi_{+}\in 2\Gamma_{E_{8}}} q^{\frac{1}{4}\pi_{+}^{2}}\left( \sum_{n~\text{even}}-\sum_{n~\text{odd}} \right) e^{-\pi\tau_{2}\left( \frac{n}{R}\right)^2 }\xrightarrow[]{R\to\infty}0.
\end{align}
Then, we see that $Z_{\mathcal{N}=0}^{(9)}$ reproduces the original partition function, i.e., the $E_{8}\times E'_{8}$ theory. 
In the limit $R\to 0$, the contributions with $n=0$ do not vanish, and we obtain the following behaviors of $Z\left(g^a,g^b \right) $:
\begin{align}
Z(1,1)&\sim 
\frac{1}{\bar{\eta}\eta^{17}}\sum_{\Pi\in \Gamma_{E_{8}}\oplus\Gamma_{E'_{8}}} q^{\frac{1}{2}\Pi^2 }\sum_{n\in\mathbb{Z}}e^{-\pi\tau_{2}\left( wR\right) ^2}
\xrightarrow[]{R\to 0}
\frac{R^{-1}}{\sqrt{\tau_{2}}}\frac{1}{\bar{\eta}\eta^{17}}\sum_{\Pi\in \Gamma_{E_{8}}\oplus\Gamma_{E'_{8}}} q^{\frac{1}{2}\Pi^2 },\\
Z(1,g)&\sim 
\frac{1}{\bar{\eta}\eta^{17}}\sum_{\pi_{+}\in 2\Gamma_{E_{8}}} q^{\frac{1}{4}\pi_{+}^{2}}\sum_{w\in\mathbb{Z}}e^{-\pi\tau_{2}\left( wR\right) ^2}\xrightarrow[]{R\to 0}\frac{R^{-1}}{\sqrt{\tau_{2}}}\frac{1}{\bar{\eta}\eta^{17}}\sum_{\pi_{+}\in 2\Gamma_{E_{8}}}q^{\frac{1}{4}\pi_{+}^{2}},\\
Z(g,1)&\sim 
\frac{1}{\bar{\eta}\eta^{17}}\sum_{\pi_{+}\in \Gamma_{E_{8}}} q^{\frac{1}{4}\pi_{+}^{2}}\sum_{w\in\mathbb{Z}}e^{-\pi\tau_{2}\left( wR\right) ^2}\xrightarrow[]{R\to 0}\frac{R^{-1}}{\sqrt{\tau_{2}}}\frac{1}{\bar{\eta}\eta^{17}}\sum_{\pi_{+}\in \Gamma_{E_{8}}}q^{\frac{1}{4}\pi_{+}^{2}},\\
Z(g,g)&\sim 
\frac{1}{\bar{\eta}\eta^{17}}\sum_{\pi_{+}\in \Gamma_{E_{8}}} e^{\pi i \pi_{+}^2}q^{\frac{1}{4}\pi_{+}^{2}}\sum_{w\in\mathbb{Z}}e^{-\pi\tau_{2}\left( wR\right) ^2}\xrightarrow[]{R\to 0}\frac{R^{-1}}{\sqrt{\tau_{2}}}\frac{1}{\bar{\eta}\eta^{17}}\sum_{\pi_{+}\in \Gamma_{E_{8}}} e^{\pi i \pi_{+}^2}q^{\frac{1}{4}\pi_{+}^{2}}.
\end{align}
We then check that $Z_{\mathcal{N}=0}^{(9)}$ in the limit $R\to 0$ provides the partition function of the $E_{8}$ theory, given by \eqref{eq: partition function of E8}.

\subsection{Spectrum}\label{subsec: spectrum non-susy CHL}
The mass formula is given in the same form as \eqref{eq: mass formula CHL string}, but the orbifold projection and the GSO projection in the twisted sector are different from those in the CHL string. As one can notice from the partition function \eqref{eq: partition function non-susy CHL}, the momentum lattice depends on the conjugacy classes of the spacetime $SO(8)$.

\subsubsection{Untwisted sector}
A surviving state in the untwisted sector takes the form
\begin{align}
\ket{\varphi}_{\text{untw}}=\frac{1}{\sqrt{2}}\left( \alpha\ket{w,n,\pi,\pi'}+(-1)^{n_1+F}g(\alpha)\ket{w,n,\pi',\pi} \right).
\end{align}
The bosonic spectrum is the same as in the untwisted sector of the CHL string, while the fermionic one is made of the combination of states with the opposite sign under the projection.

Let us first consider the case $N_{L}=1$ and $P_{L}=0$.
In the vector conjugacy class with $F=0$, we obtain the gravitational multiplet and gauge bosons of $U(1)_{r}^{d}\times U(1)_{l}^{d} \times U(1)_{L}^{8}$ from $\alpha_{-1}^{\mu}\ket{0}$ and $\alpha_{+,-1}^{I}\ket{0}$.
In the spinor conjugacy class with $F=1$, the left-moving states $\alpha_{-1}^{\mu}\ket{0}$ and $\alpha_{+,-1}^{I}\ket{0}$ are projected out, while there are eight massless spinors which come from $\alpha_{-,-1}^{I}\ket{0}$.

With $N_{L}=0$, the following massless state survives under the projection:
\begin{align}
\label{eq: roots non-susy untwsted}
\frac{1}{\sqrt{2}}\left( \ket{w,n,\pi,\pi'}+(-)^{n_{1}+F}\ket{w,n,\pi',\pi} \right),
\end{align}
where $\left( w,n,\pi,\pi' \right) $ is a solution of \eqref{eq: p_R=0} and \eqref{eq: rewritten roots untwisted CHL}. The number of bosonic and fermionic states with $P_{L}^2=1$ is always the same since $\pi\neq \pi'$, i.e, if one find a solution of \eqref{eq: p_R=0} and \eqref{eq: rewritten roots untwisted CHL}, then the following combinations always survive under the projection:
\begin{align}
\label{eq: short weight vector spinor}
\frac{1}{\sqrt{2}}\left( \ket{w,n,\pi,\pi'}+(-1)^{n_{1}}\ket{w,n,\pi',\pi} \right),~~~\text{for bosonic states},\\
\frac{1}{\sqrt{2}}\left( \ket{w,n,\pi,\pi'}-(-1)^{n_{1}}\ket{w,n,\pi',\pi} \right),~~~\text{for fermionic states}.
\end{align}
On the other hand, for $P_{L}^2=2$, the solutions must satisfy $\pi=\pi'$, and hence the degrees of freedom of bosonic and fermionic states are in general different; the solutions with $n_{1}$ even give gauge bosons with long roots, while those with $n_{1}$ odd correspond to long weights of a representation in which spinors transform. 

To summarize, the procedure to identify the massless spectrum in the untwisted sector is as follows:
\begin{enumerate}
	\item There are a gravitational multiplet, gauge bosons of $U(1)_{r}^{d}\times U(1)_{l}^{d} \times U(1)_{L}^{8}$ and eight neutral spinors in the massless spectrum at any point in the moduli space.
	\item For a given moduli $\left(E,a \right)$, find a set of solutions $\left(w,n,\pi_{+}\right)$ for \eqref{eq: p_R=0} and \eqref{eq: rewritten roots untwisted CHL}.
	\item For $P_{L}^2=1$, the solution is a short root element in both the vector and spinor conjugacy classes such as in \eqref{eq: short weight vector spinor}.
	\item For $P_{L}^2=2$, the solution is a long root (weight) element in the vector (spinor) conjugacy class if $n_{1}$ is even (odd).
\end{enumerate}
As well as in the CHL model, if $d=1$, there is no solution of $P_{L}^2=2$ with $n_{1}$ even and thus the symmetry enhancement to non-ADE algebras can not be realized. On the other hand, we can obtain solutions of $P_{L}^2=2$ with $n_{1}$ odd, and thus massless spinors can transform in a representation with long weights even if $d=1$.

\subsubsection{Twisted sector}
 In the twisted sector, the GSO projection for right-moving states is opposite to the untwisted sector. As was done in subsection \ref{subsec: massless spectrum CHL string}, we can figure out the orbifold projection for the twisted sector from the partition function. For the co-spinor conjugacy class, the left-moving partition function is the same as in the CHL model. As for the scalar conjugacy class, the partition function indicates that the projection is performed to be the opposite sign to the twisted sector in the CHL model. Then, the twisted states in the non-supersymmetric CHL model can be expressed as
\begin{align}
\label{eq: twisted states non-susy}
\ket{\varphi}_{\text{tw}}=\alpha\ket{\pi_{+},w,n}+(-1)^{s+F}g(\alpha)\ket{\pi_{+},w,n}.
\end{align}
As well as in the CHL model, all twisted states satisfying the level-matching condition survive under the projection\footnote{In the scalar conjugacy class, $N_{R}+a_{R}$ in \eqref{eq: level-matching twisted sector} must be a half-integer in order to survive under the GSO projection, and then the level-matching condition requires that $s$ be $0$ ($1$) for $N_{-}$ even (odd).}.

Let us next identify the conditions on $(P_L,p_R)$ for massless co-spinors and scalars.
\subsubsection*{Co-spinor conjugacy class}
Since the difference from the spinor conjugacy class is only the spin-chirality, $M_{R}^2=0$ is satisfied only when $N_{R}=0$ and $p_{R}=0$. The conditions on $(P_{L},p_{R})$ are hence the same as in the twisted sector of the CHL string, which are given by \eqref{eq: p_R=0} and \eqref{eq: roots twisted CHL}. 

\subsubsection*{Scalar conjugacy class}
In the scalar conjugacy class, the situation is somewhat different from the others because $M_{R}^2=0$ leads to $N_{R}=0$ and $p_{R}^2=1$, due to the GSO projection on the right-moving spectrum. Thus, the conditions on $(P_L,p_R)$ can not be written as simple as eqs.\eqref{eq: p_R=0} and \eqref{eq: rewritten roots untwisted CHL}.
Substituting the expression \eqref{eq: pR} into $p_{R}^2=1$, we find
\begin{align}
\label{eq: pR2=1 rewritten}
q  G^{-1}q^t=2,
\end{align}
where $q:=\sqrt{2}p_{R}e^t= \pi_{+} \cdot a-wE+n$. Note that the condition $n\in\mathbb{Z}^d$ can be written in terms of $q$:
\begin{align}
\label{eq: n integer scalar}
q-\pi_{+}\cdot a+wE\in \mathbb{Z}^d.
\end{align}
As for the left-movers, there are the following two possibilities of $M_{L}^2=0$:
\begin{itemize}
	\item Neutral scalars\footnote{We here use ``neutral scalar" to mean that the scalar is a singlet of the non-Abelian enhanced group. The scalar can be charged under non-enhanced $U(1)$'s.}: $N_{L}=\frac{1}{2}$, $P_{L}^2=0$;\\
	The condition $P_{L}^2=0$ (i.e. $p_{L}=\ell_{+}=0$) implies $p_{R}=\sqrt{2}we$ and $\pi_{+}=2wa$, and we can simplify the conditions on $\left(P_{L},p_{R} \right)$ as
	\begin{align}
	\label{eq: scalar N=1/2}
		\pi_{+}=2wa,~~~wGw^t=\frac{1}{2},~~~n=-wE^t.
	\end{align}
	One can show that these massless states are singlets of the non-Abelian gauge group. In order to see that, let $( \pi_{+}^{(v)}, w^{(v)},n^{(v)}) $ be a solution of \eqref{eq: p_R=0} and \eqref{eq: roots untwisted CHL}, and $( \pi_{+}^{(o)}, w^{(o)},n^{(o)}) $ be that of \eqref{eq: scalar N=1/2}. Then we check that $( \pi_{+}^{(o)}, w^{(o)},n^{(o)}) $ is orthogonal to $(\pi_{+}^{(v)}, w^{(v)},n^{(v)})$:
	\begin{align}
		\frac{1}{2}\pi_{+}^{(v)}\cdot \pi_{+}^{(o)}+w^{(v)}n^{(o)t}+n^{(v)}w^{(o)t}=\left( \pi_{+}^{(v)}\cdot a-w^{(v)}E+n^{(v)}\right) w^{(o)t}=0,
	\end{align}
	where we used $\pi_{+}^{(o)}=2w^{(o)}a$ and $n^{(o)}=-w^{(o)}E^{t}$ in the first equality.
	
	\item Charged scalars: $N_{L}=0$, $P_{L}^2=1$;\\
	The condition $P_{L}^2=1$ can be explicitly written as
	\begin{align}
	\label{eq: scalar N=0}
	\left( \pi_{+}-2wa \right)^2+\left( q +2wG \right)G^{-1}\left( q +2wG \right)^t=2.
	\end{align}
	We also notice that the norm of the internal momenta vanishes:
	\begin{align}
	P_{L}^2-p_{R}^2=\frac{1}{2}\pi_{+}^2+2wn^t=0.
	\end{align}
	In the case without Wilson lines, in particular, the condition \eqref{eq: scalar N=0} requires either $\pi_{+}=0$ or $\pi_{+}^2=2$. In the former case, we find $q=-wG$ by substituting \eqref{eq: pR2=1 rewritten} into \eqref{eq: scalar N=0}. Under $a=0$ and $q=-wG$, the conditions \eqref{eq: pR2=1 rewritten} and \eqref{eq: n integer scalar} leads to $wGw^t=2$ and $wB\in\mathbb{Z}^d$ respectively.
	In the latter case, the condition \eqref{eq: scalar N=0} yields $q=-2wG$, and then we find $2wGw^t=1$ and $-w\left(G+B \right) \in\mathbb{Z}^d$ from \eqref{eq: pR2=1 rewritten} and \eqref{eq: n integer scalar}. These conditions are nothing but \eqref{eq: scalar N=1/2} with $a=0$, and thus we obtain scalars in the adjoint representation of the $E_8$. Indeed, one can check that if the moduli satisfy $a=0$ and \eqref{eq: scalar N=1/2} for a set of $(\pi_{+};w,n)\in\Gamma_{E_{8}}\oplus (\mathbb{Z}^d+\hat{w}^1/2)\oplus \mathbb{Z}^d$, then the gauge symmetry is enhanced to $E_8$.
\end{itemize}

From the mass formula \eqref{eq: mass formula CHL string}, we see that tachyonic states appear if $N_{L}=N_{R}=0$ and the following relation is satisfied.
\begin{align}
\label{eq: condition tachyon exists}
	p_{R}^2=P_{L}^2<1,~~~~\text{for }P\in \mathbb{I}^*+V.
\end{align}
Explicitly writing, we obtain the bound on the moduli below which the theory is tachyonic:
\begin{align}
\label{eq: condition2 tachyon exists}
qG^{-1}q^t=\left( \pi_{+}-2wa \right)^2+\left( q +2wG \right)G^{-1}\left( q +2wG \right)^t<2,
\end{align}
for $(\pi_{+};w,n)\in\Gamma_{E_{8}}\oplus (\mathbb{Z}^d+\hat{w}^1/2)\oplus \mathbb{Z}^d$ that minimizes the quantity $q  G^{-1}q^t$. 
Note that the moduli must also satisfy eq.\eqref{eq: n integer scalar} for such an element $(\pi_{+};w,n)$. Assuming again $a=0$ which implies that the bound \eqref{eq: condition2 tachyon exists} holds only for $\pi_{+}=0$, it turns out that the bound \eqref{eq: condition2 tachyon exists} can be rewritten as
\begin{align}
\label{eq: condition tachyon exists a=0}
	wGw^t<2.
\end{align}
Recalling $w^1\in\mathbb{Z}+1/2$ and $w^{m\neq 1}\in\mathbb{Z}$, there are tachyonic states in the region where $G_{11}<1$ and $B_{1m}\in 2\mathbb{Z}$. Note that the constraint on $B$ comes from the condition \eqref{eq: n integer scalar} with $w=(\pm 1/2,0^{d-1})$.

\subsection{Relevance to CHL strings}\label{subsec: relevance to CHL}
Let $G_{\text{CHL}}$ and $G_{\mathcal{N}=0}$ denote the gauge groups realized at a point $(E,a)$ in the CHL model and non-supersymmetric CHL model respectively. We also denote a set of the nonzero roots of a non-Abelian group $G$ as $\Delta_{G}$. According to the result in subsection \ref{subsec: massless spectrum CHL string}, $\Delta_{G_\text{CHL}}$ is expressed as
\begin{align}
\label{eq: set of roots CHL}
\Delta_{G_\text{CHL}} =\left\lbrace P\in\Gamma^{8+d,d}\left|~\eqref{eq: p_R=0},~\eqref{eq: roots untwisted CHL} \right.  \right\rbrace + \left\lbrace P\in\mathbb{I}^*+V\left|~\eqref{eq: p_R=0},~\eqref{eq: roots twisted CHL} \right.  \right\rbrace,
\end{align}
where the first and second sets in r.h.s. come from the untwisted and twisted sectors respectively.
As we have seen in the previous subsection, in the non-supersymmetric CHL model, spacetime vectors arise only from the untwisted sector, and hence the first set in r.h.s. of \eqref{eq: set of roots CHL} corresponds to $\Delta_{G_{\mathcal{N}=0}}$. Namely, we can express $\Delta_{G_{\mathcal{N}=0}}$ as a subset of $\Delta_{G_{\text{CHL}}}$ as follows:
\begin{align}
\label{eq: Delta G non-susy}
\Delta_{G_{\mathcal{N}=0}}=\left\lbrace P\in\Delta_{G_\text{CHL}} \left|~ w^1 \in\mathbb{Z} \right. \right\rbrace,
\end{align}
It also turns out that the residual 
\begin{align}
\label{eq: residual}
\Delta_{G_{\mathcal{N}=0}}\backslash\Delta_{G_\text{CHL}}=\left\lbrace P\in\Delta_{G_\text{CHL}} \left| ~w^1 \in\mathbb{Z} +\frac{1}{2}\right.  \right\rbrace
\end{align}
gives the representation of $G_{\mathcal{N}=0}$ in which massless co-spinors transform.
Note that the number of long roots is maintained under the process of supersymmetry breaking. 

\subsection{$d=1$}\label{subsec: d=1}
As one of the simplest examples in the case $d=1$, let us consider the following moduli:
\begin{align}
\label{eq: moduli d=1}
	a=0,~~~~E=G=R^2.
\end{align}
The condition $p_{R}=0$ with no Wilson line yields
\begin{align}
\label{eq: n integer d=1}
n= wR^2\in\mathbb{Z}.
\end{align}
We next apply the procedure given in subsection \ref{subsec: spectrum non-susy CHL} to the case $d=1$ with the moduli \eqref{eq: moduli d=1} according to the spacetime $SO(8)$ conjugacy classes.

\subsubsection*{Vector}
In the vector conjugacy class, we use the condition \eqref{eq: rewritten roots untwisted CHL}. As mentioned in section \ref{sec: CHL string}, there is no solution with a long root in the case $d=1$.
For short roots, $\left(w,\pi_{+} \right) \in\mathbb{Z}\oplus\Gamma_{E_{8}}$ must satisfy
\begin{align}
\label{eq: short root d=1}
\frac{1}{2}\pi_{+}^2+2\left( wR\right) ^2=1.
\end{align}
It is obvious that this condition \eqref{eq: short root d=1} holds for either $\pi_{+}^2=2$ ($w=0$) or $\pi_{+}=0$ ($(wR)^2=1/2$). In the former case, the other condition \eqref{eq: n integer d=1} is always satisfied and we find the $240$ solutions, which correspond to the nonzero roots of $E_{8}$:
\begin{align}
\label{eq: E8 nonzero roots}
	\Delta_{E_{8}}=\left\lbrace \left(\underline{\pm,\pm,0,0,0,0,0,0}\right),\frac{1}{2}\left(\underline{\pm,\pm,\pm,\pm,\pm,\pm,\pm,\pm}_{+}\right)\right\rbrace ,
\end{align}
where the underline indicates all permutations of the slots and the subscript $+$ implies that permutations with the number of $+1$ even are kept. In the latter case, whereas, eq.\eqref{eq: n integer d=1} can not be satisfied for any $w\in\mathbb{Z}$ since inserting $(wR)^2=1/2$ into \eqref{eq: n integer d=1} yields $n=1/(2w)$. As a result, with the moduli \eqref{eq: moduli d=1}, we obtain the $E_{8}\times U(1)_{l}$ enhanced gauge symmetry.
Note that in the CHL model, $w$ can take a half-integer in the twisted sector, and the $U(1)_{l}$ is enhanced to $SU(2)$ at $R=\sqrt{2}$.

\subsubsection*{Spinor}
The conditions on short weight elements take the same form as in the vector conjugacy class, and thus we find the massless spinors in $\boldsymbol{248}$ of the $E_8$.
Unlike in the vector, due to $n$ odd, one can also find the solutions with long weights at a special radius:
\begin{align}
\label{eq: long weight d=1}
\frac{1}{2}\pi_{+}^2+2\left( wR\right) ^2=2,~~~~\left( \pi_{+}\in 2\Gamma_{E_{8}}\right).
\end{align}
This holds for only when $\pi_{+}=0$ and $(wR)^2=1$. Inserting $(wR)^2=1$ into eq. $(wR)^2=1$, we find $1/w\in \mathbb{Z}$ which means that there are $E_8$-singlet spinors with $(w,n;\pi_{+})=(\pm 1,\pm 1; 0)$ at $R=1$. 

\subsubsection*{Co-spinor}
The conditions for massless states in this class are given by the same form as eqs.\eqref{eq: n integer d=1} and \eqref{eq: short root d=1} but with $w\in\mathbb{Z}+1/2$.
The second term in r.h.s. of the condition \eqref{eq: short root d=1} can not be zero, and thus we obtain $\pi_{+}=0$ and $(wR)^2=1/2$ from eq.\eqref{eq: short root d=1}.
Since the other condition \eqref{eq: n integer d=1} with $(wR)^2=1/2$ leads to 
\begin{align}
	n=\frac{1}{w}\left(wR \right)^2=\frac{1}{2w}\in\mathbb{Z},
\end{align}
it turns out that there exist $E_{8}$-singlet co-spinors with $(w,n;\pi_{+})=(\pm 1/2,\pm 1; 0)$ at $R=\sqrt{2}$.
Note that in the CHL model, the corresponding massless states are in the vector conjugacy class and $U(1)_{l}$ is enhanced to $SU(2)$ at $R=\sqrt{2}$.

\subsubsection*{Scalar}
In this class, we use either the condition \eqref{eq: scalar N=1/2} or \eqref{eq: scalar N=0}, in addition to \eqref{eq: pR2=1 rewritten} and \eqref{eq: n integer scalar}.
Inserting the moduli \eqref{eq: moduli d=1} into the conditions \eqref{eq: pR2=1 rewritten} and \eqref{eq: n integer scalar}, we get
\begin{align}
\label{eq: conditions for scalar d=1}
	q^2=2R^2,~~~~q+wR^2\in\mathbb{Z}.
\end{align} 
The conditions \eqref{eq: scalar N=1/2} for neutral scalars are written as
\begin{align}
	\pi_{+}=0,~~~~2\left( w R\right) ^2=1,~~~~n=-wR^2=-\frac{1}{2w}\in\mathbb{Z},
\end{align}
We thus find $E_8$-singlet scalars with $(w,n;\pi_{+})=\left( \pm 1/2, \mp 1;0\right)$ at the special radius $R=\sqrt{2}$.
Recall that these massless states are excited by the oscillators $\alpha_{-,-\frac{1}{2}}^{I}$, and the degree of the degeneracy is sixteen.
As for charged scalars, the condition \eqref{eq: scalar N=0} gives
\begin{align}
\label{eq: short root scalar N=0}
	\pi_{+}^2+\frac{1}{R^2}\left(q+2wR^2 \right)^2 =2.
\end{align}
If $\pi_{+}^2=2$, then eq.\eqref{eq: short root scalar N=0} gives $q=-2wR^2$. Substituting this into eqs.\eqref{eq: conditions for scalar d=1}, we get
\begin{align}
	w^2R^2=\frac{1}{2},~~~~\frac{1}{2w}\in\mathbb{Z}.
\end{align}
At $R=\sqrt{2}$, hence, there are $E_{8}$-charged massless scalars with $(w,n;\pi_{+})=(\pm \frac{1}{2}, \mp 1;\pi_{+}\in\Delta_{E_{8}})$.
Combining with the $E_8$-neutral scalars found above at the same radius, we obtain the scalars transforming in the adjoint representation of the $E_{8}$.
On the other hand, if $\pi_{+}=0$, then we find from eq.\eqref{eq: short root scalar N=0} 
\begin{align}
\label{eq: scalar N=0, singlet}
	\left(q+2wR^2 \right)^2 = q^2+4wR^2\left( q+wR^2\right) =2R^2~\overset{q^2=2R^2}{\Longrightarrow}~q=-wR^2.
\end{align}
It is clear that the second condition in \eqref{eq: conditions for scalar d=1} is satisfied for any $w\in \mathbb{Z}+1/2$, due to $q+wR^2=0$.
The first condition in \eqref{eq: conditions for scalar d=1} with $q=-wR^2$ yields $w^2R^2=2$. 
Thus, at $\sqrt{2}R^{-1}=|w|\in\mathbb{Z}_{\geq 0}+1/2$, we obtain the $E_8$-singlet scalars with $(w,n;\pi_{+})=(\pm \sqrt{2}R^{-1},0;0)$.
Note that the maximum radius at which such $U(1)_{l}$-charged scalars appear is $R=2\sqrt{2}$. This is indeed the critical radius where the theory is tachyon-free, as we will see below.
In Table \ref{table: additional massless states d=1}, we summarize the above result of the additional massless states and the corresponding special radii.

Let us next figure out the region where tachyonic states appear. As we are now assuming $a=0$, such region is given by \eqref{eq: condition tachyon exists a=0} which is expressed as follows in the case $d=1$:
\begin{align}
	R<\frac{\sqrt{2}}{|w|}\leq 2\sqrt{2}.
\end{align}
This means that the massless neutral scalars we found above at $R=\sqrt{2}/|w|$ become tachyonic in the region $R<\sqrt{2}/|w|$.

\begin{table}[t]
	\centering
	\begin{tabular}{c|c|c|c|c}
		\hline
		& $R=1$ & \multicolumn{2}{|c|}{$R=\sqrt{2}$} & $R=\frac{\sqrt{2}}{|w|}$ ($w\in\mathbb{Z}+\frac{1}{2}$)\\ \hline
		space-time $SO(8)$ reps.&$\boldsymbol{8}_{S}$ &$\boldsymbol{8}_{C}$& scalar & scalar  \\
		$E_{8}$ reps.& singlet & singlet & adjoint & singlet \\
		$U(1)_{l}$-charge $(w,n)$& $\left(\pm 1,\pm 1 \right) $ &$\left(\pm \frac{1}{2},\pm 1 \right)$ & $\left(\pm \frac{1}{2},\mp 1 \right)$ & $\left(\pm \sqrt{2}R^{-1},0 \right) $\\\hline
	\end{tabular}
	\caption{The extra massless states appearing at the special radii in the case $d=1$ without the Wilson line.}
	\label{table: additional massless states d=1}
\end{table}

\subsection{$d=2$}\label{subsec: d=2}
The case $d=2$ is more interesting since we can see the enhancements to non-simply-laced groups. In this subsection, we will give two examples of the enhancements. 
The derivation of all solutions of the conditions for the massless spectrum is straightforward but quite tedious. So, we only give the results which are summarized in tables in appendix \ref{app: table d=2}. In this section, we focus on the points where the maximal enhancements in the eight-dimensional CHL model occur (see ref. \cite{Font:2021uyw} for more detail).

\subsubsection{Example: $C_9 \times C_1$ gauge symmetry}
As a first example, let us consider the following point in the moduli space:
\begin{align}
\label{eq: moduli1 d=2}
E=\left(\begin{array}{cc}
2&0  \\
-2&1
\end{array}
\right), ~~~a_{1}=\frac{\omega_{1}}{3},~~~a_{2}=0,
\end{align}
where
\begin{align}
\omega_{1}=-\frac{1}{2}\left(-1,1,1,1,1,1,1,-5 \right).
\end{align}
In the CHL model, at this point, $U(1)_{L}^{8}\times U(1)_{l}^2$ is enhanced to $C_{10}$ \cite{Font:2021uyw}. The condition \eqref{eq: p_R=0} with the above moduli becomes
\begin{align}
\label{eq: n integer ex1 d=2}
	n_{1}=-\pi_{+}\cdot a_1+2\left( w^1-w^2\right) \in\mathbb{Z},~~~n_{2}=w^2\in\mathbb{Z}.
\end{align}
The metric $G$ and asymmetric tensor $B$ can be extracted from $E$ and $a$:
\begin{align}
	G=\frac{1}{2}\left( E+E^t\right) -a\cdot a
	=\left(\begin{array}{cc}
	\frac{10}{9}&-1  \\
	-1&1
	\end{array}
	\right),~~~
	B=-\frac{1}{2}\left( E-E^t\right)
	=\left(\begin{array}{cc}
	0&-1  \\
	1&0
	\end{array}
	\right).
\end{align}
Inserting the metric $G$ into \eqref{eq: rewritten roots untwisted CHL}, we get the condition on short root elements
\begin{align}
\label{eq: condition short roots ex1 d=2}
	\left(\pi_{+}-2w^1 a^1 \right) +\frac{4}{9}\left( w^1\right)^2+4\left(w^1-w^2 \right) ^2=2,
\end{align}
and on long root elements
\begin{align}
\label{eq: condition long roots ex1 d=2}
\left(\pi'_{+}-w^1 a^1 \right) +\frac{1}{9}\left( w^1\right)^2+\left(w^1-w^2 \right) ^2=1.
\end{align}
The condition \eqref{eq: pR2=1 rewritten} in the scalar conjugacy class is written as
\begin{align}
\label{eq: d=2 ex1 scalar}
	9\left( q_{1}+q_{2} \right)^2+q_{2}^2=\tilde{q}_{1}^{2}+q_{2}^2=2,
\end{align}
where $q_1=\pi_{+}\cdot a_1-2w^1+2w^2+n_1$ and $q_2=-w^2+n_2$, and we define $\tilde{q}_{1}=3q_1+3q_2$.
Note that both $\tilde{q}_1$ and $q_2$ are integers due to $3a_1\in\Gamma_{E_{8}}$, and thus we find the four solutions of eq.\eqref{eq: d=2 ex1 scalar}
\begin{align}
\label{eq: tilde q1 q2}
	\left(\tilde{q}_1,q_2\right)=\pm\left(1,1 \right) , \pm\left(1,-1 \right).
\end{align}
We can express the condition \eqref{eq: scalar N=0} for charged scalars as
\begin{align}
	\left(\Pi_{+}-2W^1a_1 \right) ^2+\frac{4}{9}\left( W^1\right)^2+4\left(W^1-W^2 \right) ^2=2.
\end{align}
Here, we define
\begin{align}
\label{eq: condition charged scalars d=2}
	\Pi_{+}:=\pi_{+}+3\tilde{q}_1a_1,~~~W^1:=w^1+\frac{3}{2}\tilde{q}_1,~~~W^2:=w^2+\frac{1}{2}\left( 3\tilde{q}_1-q_2\right) .
\end{align}
Since $\tilde{q}_1$ and $q_2$ are restricted as in eq.\eqref{eq: tilde q1 q2} and $w^1\in\mathbb{Z}+1/2$, we notice $\Pi_{+}\in\Gamma_{E_{8}}$, $W^1\in\mathbb{Z}$ and $W^2\in\mathbb{Z}$.
The condition \eqref{eq: n integer scalar} can be also written as
\begin{align}
\label{eq: n integer scalar 2}
	n_1=N_1+3\tilde{q}_1\in\mathbb{Z},~~~~n_2=N_2-\frac{1}{2}\left( 3\tilde{q}_1-q_2\right) \in\mathbb{Z},
\end{align}
where we define
\begin{align}
	N_1:=-\Pi_{+}\cdot a_1+2\left( W^1-W^2\right),~~~N_2:=W^2.
\end{align}
The condition \eqref{eq: n integer scalar 2} clearly requires that $N_1$ and $N_2$ be integers. Therefore it turns out that the solutions $(W^1,W^2;N_1,N_2;\Pi_{+})$ of \eqref{eq: condition charged scalars d=2} and \eqref{eq: n integer scalar 2} are nothing but those of \eqref{eq: condition short roots ex1 d=2} and \eqref{eq: n integer ex1 d=2}.
Namely, the elements $(w^1,w^2;n_1,n_2;\pi_{+})$ on which charged massless scalars lie are given by the short root solutions in the vector (or spinor) conjugacy class shifted by the vector
\begin{align}
\label{eq: shift vector scalar}
\left( -\frac{3}{2}\tilde{q}_1,-\frac{1}{2}(\tilde{q}_1+q_2);3\tilde{q}_1,-\frac{1}{2}(3\tilde{q}_1-q_2);-3\tilde{q}_1a_1\right).
\end{align}
Furthermore, one can find that the shift vector \eqref{eq: shift vector scalar} is a solution of the conditions for neutral scalars given by \eqref{eq: scalar N=0}. Since there are four independent choices of $(\tilde{q}_1,q_2)$ given by \eqref{eq: tilde q1 q2}, we obtain the four copies of massless scalars transforming in the representation which consists of a set of elements satisfying the conditions on short root and eight neutral elements.

The result of the solutions is summarized in table \ref{table: C9 x C1 weights}. In the vector conjugacy class, the solutions correspond to the nonzero roots of $C_{1} \times C_{9}$. There is no long weight element in the spinor conjugacy class, and combining the short weight solutions with eight neutral elements, we obtain the massless spinors transforming in $\boldsymbol{152}\oplus\boldsymbol{1}$ of the $C_{9}$. We also find the co-spinors in $\left( \boldsymbol{2},\boldsymbol{18}\right) $ of the $C_{1}\times C_{9}$ in the twisted sector. It is natural for the co-spinors to transform in the bi-fundamental representation because $G_{\text{CHL}}=C_{10}$, $G_{\mathcal{N}=0}=C_{9}\times C_1$ and the residual $\Delta_{C_9\times C_1}\backslash\Delta_{C_{10}}$ gives a set of weight vectors associated to the co-spinors, as mentioned in subsection \ref{subsec: relevance to CHL}. 
As we have seen above, there are the four copies of massless states transforming in $\boldsymbol{152}\oplus\boldsymbol{1}$ of the $C_{9}$ in the scalar conjugacy class.

\subsubsection{Example: $A_4 \times C_2 \times C_4$ gauge symmetry}
As a second example, let us consider the moduli
\begin{align}
\label{eq: moduli2 d=2}
E=\left(\begin{array}{cc}
2&0  \\
0&1
\end{array}
\right), ~~~a_{1}=\frac{\omega_{6}}{2},~~~a_{2}=\frac{\omega_{3}}{5},
\end{align}
where
\begin{align}
\omega_{6}=\left(0,0,0,0,0,0,-1,1 \right) ,~~~~\omega_{3}=\left(0,0,0,-1,-1,-1,-1,4 \right).
\end{align}
In the CHL model with this background, the gauge symmetry is enhanced to $A_{4}\times C_{6}$ \cite{Font:2021uyw}. 
Inserting the above $E$ into eq.\eqref{eq: p_R=0}, we get
\begin{subequations}
	\label{eq: n integer ex2 d=2}
\begin{align}
	n_{1}&=-\pi_{+}\cdot a_{1}+ 2w^1\in\mathbb{Z},\\
	n_{2}&=-\pi_{+}\cdot a_{2}+ w^2\in\mathbb{Z}.
\end{align}
\end{subequations}
Obviously $B=0$, and the metric is 
\begin{align}
\label{eq: metric2 d=2}
G=E-a\cdot a
=\left(\begin{array}{cc}
\frac{3}{2}&-\frac{1}{2}  \\
-\frac{1}{2} &\frac{1}{5}
\end{array}
\right).
\end{align}
Using this metric $G$, the condition on short root elements is expressed as
\begin{align}
\label{eq: condition short roots ex2 d=2}
\left( \pi_{+}-2w^i a_i\right) ^{2}+\left(w^1\right)^2+\frac{1}{5}\left(5w^1 - 2w^2\right)^2   = 2.
\end{align}
As for long roots, using $\pi'_{+}=\pi_{+}/2\in\Gamma_{E_{8}}$, we obtain
\begin{align}
\label{eq: condition long roots ex2 d=2}
\left( \pi'_{+}-w^i a_i\right) ^{2}+\frac{1}{4}\left(w^1\right)^2+\frac{1}{20}\left(5w^1 - 2w^2\right)^2   = 1.
\end{align}
The condition \eqref{eq: pR2=1 rewritten} in the scalar conjugacy class is
\begin{align}
\label{eq: d=2 ex2 scalar}
	\left( 2q_1+5q_2 \right)^2+5q_2^2=\left( 2q_1+5q_2 \right)^2+\frac{1}{5}\left( 5q_2\right) ^2=2,
\end{align}
where $q_1=\pi_{+}\cdot a_1-2w^1+n_1$ and $q_2=\pi_{+}\cdot a_2-w^2+n_2$. Since $2q_1$ and $5q_2$ are integers (note $2a_1\in\Gamma_{E_{8}}$ and $5a_2\in\Gamma_{E_{8}}$), there is no a pair $\left(2q_1,5q_2 \right) $ of integers satisfying eq.\eqref{eq: d=2 ex2 scalar}, i.e. no scalars in the massless spectrum. 

All the solutions with the moduli \eqref{eq: moduli2 d=2} are given in table \ref{table: A4 x C4 x C2 weights}. We identify the solutions in the vector conjugacy class as the nonzero roots of $A_4\times C_{2}\times C_{4}$. As expected by \eqref{eq: residual} with $G_{\text{CHL}}=A_{4}\times C_{6}$ and $G_{\mathcal{N}=0}=A_{4}\times C_{2}\times C_{4}$, the massless co-spinors transform in $\left(\boldsymbol{4},\boldsymbol{8}\right) $ of the $C_{2}\times C_{4}$.
In this example, unlike the previous one, there are solutions for long weights with $n_1$ odd, and we obtain the spinors in $\left( \boldsymbol{24},\boldsymbol{1},\boldsymbol{1}\right)\oplus\left( \boldsymbol{1},\boldsymbol{5}\oplus\boldsymbol{1},\boldsymbol{1}\right) \oplus 
\left( \boldsymbol{1},\boldsymbol{1},\boldsymbol{42}\oplus\boldsymbol{1}\oplus\boldsymbol{1}\right) $ of the $A_{4}\times C_{2}\times C_{4}$. 

\section{Cosmological constant}\label{sec: cosmological const}
In this section, we evaluate the leading behavior of the cosmological constant in the region where supersymmetry is asymptotically restored. We will see that as in \eqref{eq: cc nf-nb}, the leading contribution is proportional to $\left( n_F-n_B\right)$ under a constraint on the Wilson line $a_{1}$. The derivation is almost the same as in other interpolating models (e.g., see section IV in ref. \cite{Abel:2015oxa} or appendix in ref. \cite{Itoyama:2020ifw}).

The one-loop cosmological constant is expressed as the integral of the torus partition function over the fundamental domain $\mathcal{F}$ of the modular group:
\begin{align}
\label{eq: cosmological const}
\Lambda^{(10-d)}=-\frac{1}{2}\left( 4\pi^2 \right)^{-\frac{10-d}{2}}\int_{\mathcal{F}}\frac{d^2 \tau}{\tau_{2}^2}Z_{\mathcal{N}=0}^{(10-d)},
\end{align}
In order to make the behavior of the supersymmetry breaking scale $m_{3/2}$ clear, we restrict our attention to the following background:
\begin{align}
\label{eq: moduli ss direction}
e^{1}_{m}=R\delta^{1}_{m},~~~e^{a}_{1}=R\delta^{m}_{1},~~~B_{m1}=0.
\end{align}
These choices of the veirbeil $e^{a}_{m}$ implies that $T^{d}$ is decomposed into $S^1\times {T^{d-1}}'$, i.e. $G_{m1}=R^2\delta^{1}_{m}$. For states with $w^1=0$, which are dominant in the region we are focusing on, the internal momenta can be written as
\begin{subequations}
	\begin{align}
	p_{L,1}&=\frac{1}{\sqrt{2}R}\left(\pi_{+} \cdot a_{1}-w'a'\cdot a_{1}+n_{1}  \right),\\
	p_{R,1}&=\frac{1}{\sqrt{2}R}\left(\pi_{+} \cdot a_{1}-w'a'\cdot a_{1}+n_{1}  \right),
	\end{align}
\end{subequations}
and 
\begin{align}
\left( P'_{L}; p'_{R}\right) =\left( \pi_{+},w',n'\right) \mathcal{E}^{(d-1)}\left( e',B',a'\right),
\end{align}
where $e'=e^{a'}_{m'}$, $B'=B_{m'n'}$, $a'=a_{m'}^{I}$ and $\left( w',n'\right) =(w^{m'},n_{m'}) $ with $m',n',a'=2,\cdots,d$. Note that states with $w^1=0$ only exist in the untwisted sector, and the massless conditions for those states are given by \eqref{eq: rewritten roots untwisted CHL} and \eqref{eq: p_R=0} with $(w,n,E,a)$ replaced by $(w',n',E',a')$, in addition to $p_{R,1}=0$:
\begin{align}
\label{eq: n=rho.a1}
	n_{1}=-\left( \pi_{+}-w'a'\right) \cdot a_{1}\in\mathbb{Z}.
\end{align}
For long roots, in particular, recall that whether the state is bosonic or fermionic depends on whether $n_{1}$ is even or odd.

Let us compute $\Lambda^{(10-d)}$ in the background \eqref{eq: moduli ss direction} with $R$ large. In this regime, the contributions with $w^1\neq 0$ are suppressed at least by the factor of $ e^{-R^2}$. Inserting the partition function \eqref{eq: partition function non-susy CHL} into the cosmological constant \eqref{eq: cosmological const} and using the Jacobi abstruse identity $V_8-S_8=0$, we find
\begin{align}
\label{eq: cc w1=0}
\Lambda^{(10-d)}&\sim-\frac{1}{2}\left( 4\pi^2  \right)^{-\frac{10-d}{2}}
\int_{\mathcal{F}}\frac{d^2 \tau}{\tau_{2}^2}Z_{B}^{(8-d)}\bar{V}_{8}\left. Z(1,g)\right| _{w^1=0}\nonumber\\
&\sim-\frac{1}{2}\left( 4\pi^2 \right)^{-\frac{10-d}{2}}
\int_{\mathcal{F}}\frac{d^2 \tau}{\tau_{2}^{\frac{12-d}{2}}}\bar{V}_{8}\frac{1}{\bar{\eta}^8\eta^{24}}\left( \frac{2\eta^3}{\vartheta_{2}}\right)^{4} \left. \sum_{\left(P_{L};p_R \right)\in \mathbb{I}}q^{\frac{1}{2}P_{L}^2 }\bar{q}^{\frac{1}{2}p_{R}^2}e^{\pi i n_1}\right|_{w^1=0}.
\end{align}
The sum over the invariant lattice $\mathbb{I}$ is expressed in terms of the theta functions:
\begin{align}
\label{eq: momentum sum 1}
\left. \sum_{\left(P_{L};p_R \right)\in I }q^{\frac{1}{2}P_{L}^2 }\bar{q}^{\frac{1}{2}p_{R}^2}e^{\pi i n_1}\right|_{w^1=0}&
=\sum_{\substack{\pi_{+}\in 2\Gamma_{E_{8}}\\w',n'\in\mathbb{Z}}}q^{\frac{1}{2}P'^{2}_{L} }\bar{q}^{\frac{1}{2}p'^{2}_{R}}\sum_{n_1\in\mathbb{Z}}\left\lbrace 
e^{\frac{4\pi \tau_{2}}{R^2}\left( n_{1}+\frac{\rho\cdot a_{1}}{2}\right)^2 }-e^{\frac{4\pi \tau_{2}}{R^2}\left( n_{1}+\frac{\rho\cdot a_{1}+1}{2}\right)^2 }\right\rbrace\nonumber\\
&=\sum_{\substack{\pi_{+}\in 2\Gamma_{E_{8}}\\w',n'\in\mathbb{Z}}}q^{\frac{1}{2}P'^{2}_{L} }\bar{q}^{\frac{1}{2}p'^{2}_{R}}\left\lbrace 
\vartheta
\begin{bmatrix} 
\frac{\rho\cdot a_{1}}{2}\\ 
0\\ 
\end{bmatrix}\left( \frac{4i\tau_{2}}{R^{2}}\right)-
\vartheta
\begin{bmatrix} 
\frac{\rho\cdot a_{1}+1}{2}\\ 
0\\ 
\end{bmatrix}\left( \frac{4i\tau_{2}}{R^{2}}\right)
\right\rbrace,
\end{align}
where we define $\rho:=\pi_{+}-w'a'$.
By using the $S$-transformation law \eqref{eq: S-trans of eta theta} of the theta functions, eq.\eqref{eq: momentum sum 1} can be rewritten as
\begin{align}
\label{eq: momentum sum 2}
&\frac{2R}{\sqrt{\tau_{2}}}\sum_{\pi_{+}\in 2\Gamma_{E_{8}}}\sum_{w',n'\in\mathbb{Z}^{d-1}}q^{\frac{1}{2}P'^{2}_{L} }\bar{q}^{\frac{1}{2}p'^{2}_{R}}
\left\lbrace 
\vartheta
\begin{bmatrix} 
0\\ 
\frac{\rho\cdot a_{1}}{2}\\ 
\end{bmatrix}\left( \frac{iR^{2}}{4\tau_{2}}\right)-
\vartheta
\begin{bmatrix} 
0\\ 
\frac{\rho\cdot a_{1}+1}{2}\\ 
\end{bmatrix}\left( \frac{iR^{2}}{4\tau_{2}}\right)
\right\rbrace\nonumber\\
&=\frac{2R}{\sqrt{\tau_{2}}}
\sum_{\pi_{+}\in 2\Gamma_{E_{8}}}\sum_{w',n'\in\mathbb{Z}^{d-1}}q^{\frac{1}{2}P'^{2}_{L} }\bar{q}^{\frac{1}{2}p'^{2}_{R}}
\sum_{n\geq 1} \cos\left[\pi (2n-1) \rho\cdot a _{1} \right]e^{-\frac{\pi (2n-1)^2R^2}{4\tau_{2}}} .
\end{align}
Substituting eq.\eqref{eq: momentum sum 2} into eq.\eqref{eq: cc w1=0}, we find
\begin{align}
\Lambda^{(10-d)}
\sim-\frac{R}{\left( 4\pi^2 \right)^{\frac{10-d}{2}}}\sum_{N_{+},N_{-}}a_{N_{+},N_{-}}
&\sum_{\pi_{+}\in 2\Gamma_{E_{8}}}\sum_{w',n'\in\mathbb{Z}^{d-1}}
\sum_{n\geq 1} \cos\left[\pi (2n-1) \rho\cdot a_{1} \right]\nonumber\\
&\times\int_{\mathcal{F}}\frac{d^2 \tau}{\tau_{2}^{\frac{13-d}{2}}}e^{2\pi i \tau_{1}M'^{2}_{-} }e^{-\pi\left( 2\tau_{2} M'^{2}_{+}+ \frac{ (2n-1)^2R^2}{4\tau_{2}}\right) },
\end{align}
where we expand the contribution from the oscillators as
\begin{align}
\label{eq: q-expansion oscillators}
\bar{V}_{8}\frac{1}{\bar{\eta}^8\eta^{24}}\left( \frac{2\eta^3}{\vartheta_{2}}\right)^{4} =\sum_{N,\tilde{N}} \hat{a}_{N,\tilde{N}}q^{N}\bar{q}^{\tilde{N}}=\sum_{N_{+},N_{-}} a_{N_{+},N_{-}}e^{2\pi i \tau_{1}N_{-}} e^{-2\pi  \tau_{2}N_{+}},
\end{align}
and define as
\begin{align}
M'^{2}_{+}=N_{+}+\frac{1}{2}\left( P'^{2}_{L}+p'^{2}_{R}\right),~~~~M'^{2}_{-}=N_{-}+\frac{1}{2}\left( P'^{2}_{L}-p'^{2}_{R}\right)=N_{-}+\frac{1}{4}\pi_{+}^2+w'n'.
\end{align}
In order to evaluate the integration, let us decompose the fundamental domain $\mathcal{F}$ into $\mathcal{F}_{\geq 1}=\left. \mathcal{F}\right| _{\tau_{2}\geq 1}$ and $\mathcal{F}_{< 1}=\left. \mathcal{F}\right| _{\tau_{2}< 1}$. For the integration over $\mathcal{F}_{< 1}$, the domain itself is finite and the integrand itself is non-singular.
Then, we can bound the integration as
\begin{align}
\int_{\mathcal{F}_{<1}}\frac{d^2 \tau}{\tau_{2}^{\frac{13-d}{2}}}e^{2\pi i \tau_{1}M'^{2}_{-} }e^{-\pi\left( 2\tau_{2} M'^{2}_{+}+ \frac{ (2n-1)^2R^2}{4\tau_{2}}\right) }<
e^{-\frac{ \pi (2n-1)^2R^2}{4} }
\int_{\mathcal{F}_{<1}}\frac{d^2 \tau}{\tau_{2}^\frac{13-d}{2}}e^{2\pi i \tau_{1}M'^{2}_{-} }
e^{-2\pi \tau_{2} M'^{2}_{+}}.
\end{align}
So, this contribution is suppressed at least by the factor $e^{-\pi R^2/4}$. 
As for $\mathcal{F}_{\geq 1}$, the integration over $\tau_{1}$ gives the level-matching(-like) condition $M'^{2}_{-}=0$. By the arithmetic-geometric mean, one can find the bound of the $\tau_{2}$-dependent factor:
\begin{align}
e^{-\pi\left( 2\tau_{2} M'^{2}_{+}+ \frac{ (2n-1)^2R^2}{4\tau_{2}}\right) }\leq
e^{-\sqrt{2}\pi (2n-1)M'_{+}R}.
\end{align}
The contribution from $\mathcal{F}_{\geq 1}$ is suppressed at least by the factor $e^{-\sqrt{2}\pi M'_{+}R}$ unless $M'^{2}_{+}=0$. Up to the exponentially suppressed terms, thus, the leading contribution is
\begin{align}
\Lambda^{(10-d)}
&\sim-\frac{R}{\left( 4\pi^2  \right)^{\frac{10-d}{2}}}\sum_{n\geq 1}
\int_{1}^{\infty}\frac{d\tau_{2}}{\tau_{2}^\frac{13-d}{2}}e^{-\frac{\pi (2n-1)^2}{4\tau_{2}}R^2}
\sum_{\left\lbrace N_{\pm},P'_{L},p'_{R} \right\rbrace \in \mathbb{L}_{0}}a_{N_{+},N_{-}}
\cos\left[\pi (2n-1) \rho\cdot a_{1}  \right]\nonumber\\
&\sim - \frac{2\Gamma\left( \frac{11-d}{2}\right) }{ \sqrt{\pi}\left( \pi^{\frac{3}{2}}R \right)^{10-d} }
\sum_{n\geq 1}(2n-1)^{d-11}
\sum_{\left\lbrace N_{\pm},P'_{L},p'_{R} \right\rbrace \in \mathbb{L}_{0}}a_{N_{+},N_{-}}
\cos\left[\pi (2n-1) \rho\cdot a_{1} \right],
\end{align}
where we define
\begin{align}
\mathbb{L}_{0}=\left\lbrace  N_{\pm},P'_{L},p'_{R} \left|~M'^{2}_{+} =M'^{2}_{-}=0 \right.  \right\rbrace.
\end{align}
Note that the constraint in $\mathbb{L}_{0}$ can be rewritten as
\begin{align}
\label{eq: constraint in L0}
N+\frac{1}{2}P'^{2}_{L}=\tilde{N}+\frac{1}{2}p'^{2}_{R}=0,
\end{align}
and $N$, $\tilde{N}$ and $\tilde{a}_{N,\tilde{N}}$ can be identified by expanding eqs.\eqref{eq: q-expansion oscillators} as
\begin{align}
\frac{1}{\eta^{24}}\left( 2\frac{\eta^3}{\vartheta_{2}}\right)^{4}&=q^{-1}+8+\mathcal{O}(q),~~~~\frac{\bar{V}_{8}}{\bar{\eta}^8}=8+\mathcal{O}(\bar{q}).
\end{align}
In order to satisfy \eqref{eq: constraint in L0}, $\tilde{N}=p'_{R}=0$ and either $N=P'_{L}=0$ or $N=-1$, $P'^{2}_{L}=2$, and we find
\begin{align}
\label{eq: leading cc cos}
\Lambda^{(10-d)}
&\sim - \frac{2\Gamma\left( \frac{11-d}{2}\right) }{ \sqrt{\pi}\left( \pi^{\frac{3}{2}}R \right)^{10-d} }
\sum_{n\geq 1}(2n-1)^{d-11}
\left( 64+8 \sum_{\rho\in \mathbb{L}'_{0}} \cos\left[\pi (2n-1) \rho\cdot a_{1} \right] \right),
\end{align}
where the first term $64$ in the parentheses comes from $\hat{a}_{0,0}$, and we define
\begin{align}
	\mathbb{L}'_{0}=\left\lbrace \rho=\pi_{+}-w'a' \left|~P'^{2}_{L}=2,~p'_{R}=0\right.  \right\rbrace.
\end{align}

Let us henceforth assume that $a_{1}$ is chosen such that $\rho\cdot a_{1}$ is an integer for all solutions of $p'_{R}=0,~P'^{2}_{L}=2$ for given $E'$ and $a'$. Since we focus on non-winding states along $x^1$, it turns out from \eqref{eq: n=rho.a1} that the integer value $\rho\cdot a_{1}$ is nothing but $n_{1}$. 
The cosine factor in r.h.s. of \eqref{eq: leading cc cos} gives $+1$ ($-1$) with $n_1$ even (odd), and the second term in the parentheses is indeed the net number of massless bosonic and fermionic states with long roots.
As mentioned in section \ref{sec: non-susy CHL string}, the net number of massless bosons and fermions with short roots is always zero, and the factor $64$ in the parentheses in \eqref{eq: leading cc cos} agrees with the degrees of freedom in the gravitational sector. Therefore, the cosmological constant \eqref{eq: leading cc cos} can be interpreted as
\begin{align}
\label{eq: leading cc nf-nb}
\Lambda^{(10-d)}
\sim \frac{\Gamma\left( \frac{11-d}{2}\right) }{ \sqrt{\pi}\left( \pi^{\frac{3}{2}}R \right)^{10-d} }
2^{d-10}\zeta\left( 11-d,\frac{1}{2}\right)\left( n_{F}-n_{B} \right),
\end{align}
where $n_{F}$ and $n_{B}$ are the degrees of freedom of massless fermionic states and bosonic states respectively. Note again that we restrict our attention to the region where $R$ is sufficiently larger than the string length scale.

One can easily see that the cosmological constant can not be suppressed with $d=1$.
In the case $d=1$, there is no solutions of $P'^{2}_{L}=2$ since $P'_{L}=\frac{1}{\sqrt{2}}\pi_{+}=\frac{1}{\sqrt{2}}\rho\in\sqrt{2}\Gamma_{E_{8}}$. Then, the leading term is independent of the Wilson line $a_1$ and always negative:
\begin{align}
\Lambda^{(9)}
\sim-\frac{24}{\pi^{14}R^9 }2^{-9}\zeta\left( 10,\frac{1}{2}\right) 64,
\end{align}
where the Hurwitz zeta function $\zeta(s,a):=\sum_{n=0}^{\infty}\left(n+a \right)^{-s}$.

In the following subsections, we will give explicit examples of the points at which the cosmological constant is exponentially suppressed in the cases $d=2$ and $d=3$.

\subsection{Suppression of $\Lambda^{(8)}$}
\label{eq: suppression d=1}
In the case $d=2$, the background is $S^{1}\times {S^{1}}'$ where the Scherk-Schwarz mechanism is implemented on $S^{1}$ and the radius $R$ of $S^1$ is supposed to be large. Let us consider the following special radius and Wilson line for ${S^{1}}'$:
\begin{align}
\label{eq: moduli for S1'}
	R'=\frac{1}{2},~~~~a'=\left( 0,0,0,0,-\frac{1}{4},-\frac{1}{4},-\frac{1}{4},\frac{3}{4} \right),~~~~E'=R'^{2}+a'\cdot a'=1.
\end{align}
Inserting the moduli \eqref{eq: moduli for S1'} into $P'^{2}_{L}=2$ and $p'_{R}=0$, we get
\begin{align}
	\left( \pi'_{+}-w' a' \right)^2 + \frac{1}{4}w'^{2}=1,~~~n'=-2\pi'_{+}\cdot a'+w'\in\mathbb{Z},
\end{align}
where $\pi'_{+}=\pi_{+}/2\in\Gamma_{E_{8}}$.
There are eight solutions to the above equations:
\begin{align}
\label{eq: solutions P'2=2 d=1}
	\left(w';n' ;\pi_{+}\right)=\pm\left(1;1;0,0,0,0,0,0,0,0\right),\pm\left(1;-1;0,0,0,0,\underline{-2,0,0},2\right).
\end{align}
In order to realize $n_{F}-n_{B}=0$, $a_{1}$ must be chosen such that $\rho\cdot a_{1}=(\pi_{+}-w'a')\cdot a_1$ is odd for all of the solutions \eqref{eq: solutions P'2=2 d=1}.
Assuming that $a_{1}$ is restricted to be an element in $\Gamma_{E_{8}}$ (i.e. $\rho\cdot a_{1}=-w'a'\cdot a_{1}$ mod $2$), one can easily find $a_{1}$ satisfying the requirement, e.g, $a_{1}=(0,0,0,0,0,0,-1,1)$. With those moduli, by following the procedure in subsection \ref{subsec: spectrum non-susy CHL}, one can find that the enhanced non-Abelian gauge symmetry is $D_{4}\times D_{5}$ and massless chiral fermions transform in $(\boldsymbol{1},\boldsymbol{45})\oplus(\boldsymbol{35}\oplus\boldsymbol{1},\boldsymbol{1})$ of the $D_{4}\times D_{5}$.

\subsection{Suppression of $\Lambda^{(7)}$}
\label{subsec: suppression d=3}
Let us turn to the case $d=3$, i.e., the background $S^1\times {T^2}'$.
We suppose the moduli for ${T^2}'$ to be chosen as
\begin{align}
\label{eq: moduli for T2'}
	E'=\left(\begin{array}{cc}
	2&0  \\
	0&1
	\end{array}
	\right), ~~~a_{2}=0,~~~a_{3}=\frac{1}{8}\left(1,1,1,1,1,1,1,-7\right).
\end{align}
In the CHL model on $T^2$ with the above moduli, the gauge group is enhanced to $A_1\times A_1 \times C_8$ \cite{Font:2021uyw}.
The conditions $P'^2_{L}=2$ and $p'_{R}=0$ with the moduli \eqref{eq: moduli for T2'} give
\begin{align}
	\left( \pi'_{+}-w^3 a_3\right)^2+2\left(w^2\right) ^2+\frac{1}{8} \left(w^3 \right) ^2=1,
\end{align}
and
\begin{align}
	n_{2}=2w^2\in\mathbb{Z},~~~n_{3}=-2\pi'_{+}\cdot a_{3}+w^3\in\mathbb{Z}.
\end{align}
Solving the above equations, we get the sixteen solutions $(w';\pi_{+})$\footnote{We omit $n'$ which does not matter to check whether the suppression is realized or not.}:
\begin{align}
\label{eq: a1 nf-nb=0 d=2}
	\pm\left(0,1;0,0,0,0,0,0,0\right),~\pm\left(0,1;\underline{2,0,0,0,0,0,0},-2\right).
\end{align}
One can check that $n_F-n_B=0$ is achieved if $a_1$ is chosen as\footnote{Of course, this $a_1$ is not the unique choice to realize $n_F-n_B=0$. For instance, if one permutates the first seven components of \eqref{eq: a1 nf-nb=0 d=2}, $n_F-n_B$ remains vanishing.}
\begin{align}
\label{eq: a1 d=3}
	a_{1}=\frac{1}{2}\left(0,0,0,0,0,0,7,1\right).
\end{align}
Note that $a_3\cdot a_1=0$ and then $\rho\cdot a_1=\pi_{+}\cdot a_1$. So, we can divide the sixteen solutions into four and twelve depending on $n_1$ being even or odd:
\begin{subequations}
	\label{eq: d=3 n1 even odd}
	\begin{align}
	&n_{1}\text{ even}:~\pm\left(0,1;0,0,0,0,0,0,0\right),~\pm\left(0,1;0,0,0,0,0,0,2,-2\right),\\
	&n_{1}\text{ odd}:~\pm\left(0,1;\underline{2,0,0,0,0,0},0,-2\right).
	\end{align}
\end{subequations}
 Thus, the second term in the parentheses in eq.\eqref{eq: leading cc cos} is $-64$ and the leading contribution of $\Lambda^{(7)}$ is canceled out.
 By following the procedure in subsection \ref{subsec: spectrum non-susy CHL}, we find that at the point \eqref{eq: moduli for T2'}, \eqref{eq: a1 d=3} and with large $R$, the enhanced non-Abelian gauge group is $A_1\times C_2\times D_6$ and there exist chiral fermions transforming in $(\boldsymbol{3},\boldsymbol{1},\boldsymbol{1})\oplus (\boldsymbol{1},\boldsymbol{5}\oplus\boldsymbol{1},\boldsymbol{1})\oplus  (\boldsymbol{1},\boldsymbol{1},\boldsymbol{77}\oplus\boldsymbol{1})$ of the $A_1\times C_2\times D_6$.

\section{Summary and discussion}\label{sec: summary}
In this paper, we have studied the reduced rank non-supersymmetric heterotic string model constructed by the asymmetric orbifold with the $\mathbb{Z}_2$-twist $\R\T(-1)^F$. We first reviewed the CHL string in which all sixteen supercharges are preserved and the rank of gauge groups is reduced to $8+d$. We then considered a natural extension of the CHL model to a non-supersymmetric one by combining the orbifold twist with a $2\pi$ spatial rotation. The conditions on the internal momenta of massless states have been figured out.
We gave concrete examples of the symmetry enhancements and the matter spectra in the cases $d=1,2$. In $d=2$, in particular, the enhanced groups are non-simply-laced and the fermions (and scalars) that non-trivially transform under the gauge group are found. We also evaluated the cosmological constant in the regime with $R$ large and argued that its exponential suppression is possible if $d\geq 2$. It turned out that in order to realize the suppression, the existence of massless fermions with long weight elements of gauge groups is crucial.

Although the conditions that internal momenta of massless states should satisfy are given in this paper, it is not shown how to solve them in an efficient way. In refs. \cite{Font:2020rsk,Font:2021uyw}, the algorithms to systematically explore the maximal enhancement of gauge groups have been proposed in heterotic strings with sixteen supercharges. By using the relations \eqref{eq: Delta G non-susy} and \eqref{eq: residual}, it is expected that the methods in refs. \cite{Font:2020rsk,Font:2021uyw} can be applied to the non-supersymmetric case, and patterns of the maximal enhancements and the corresponding representations of co-spinors can be figured out. However, it is unclear whether it can be also done in the spinor and scalar conjugacy classes because the orbifold projection in these classes is opposite to the CHL strings and the momentum lattices do not have simple relevance, such as \eqref{eq: Delta G non-susy} and \eqref{eq: residual}.

\subsection{Moduli stabilization}
In this paper, we did not take the moduli stabilization into consideration. In general, some moduli in non-supersymmetric string theory are frozen because they acquire potentials at the quantum level. For instance, we have already obtained the one-loop effective potential of the Wilson line moduli by eq.\eqref{eq: leading cc cos}, and we can perform the stability analysis of the Wilson lines. 
Let us again assume that the Wilson line $a_{1}$ satisfy $\rho\cdot a_1=n_1\in\mathbb{Z}$ for all solutions of $p'_{R}=0$ and $P'^{2}_{R}=2$. Note that such points are critical points for the $a_{1}$-directions:
\begin{align}
	\frac{\partial \Lambda^{(10-d)}}{\partial a_{1}^{I}}\sim \sum_{\rho}\rho^{I}\sin \left[\pi (2n-1) \rho\cdot a_{1} \right] =0.
\end{align}
Let $\Delta_{+}$ and $\Delta_{-}$ be sets of $\rho$ satisfying $p'_{R}=0$ and $P'^{2}_{R}=2$ with $\rho\cdot a_1$ even and odd respectively:
\begin{align}
	\Delta_{+}=\left\lbrace \rho \left|~ p'_{R}=0,~P'^{2}_{R}=2,~\rho\cdot a_{1}\in 2\mathbb{Z} \right.  \right\rbrace,~~~~ \Delta_{-}=\left\lbrace \rho \left|~ p'_{R}=0,~P'^{2}_{R}=2,~\rho\cdot a_{1}\in 2\mathbb{Z}+1 \right.  \right\rbrace.
\end{align}
In order for the cosmological constant \eqref{eq: leading cc cos} to vanish, the difference between the numbers of elements in $\Delta_{+}$ and $\Delta_{-}$ must be $-8$:
\begin{align}\label{eq: condition nF-nB=0}
|\Delta_{+}|-|\Delta_{-}|=-8.
\end{align}
The Hessian $H_{IJ}$ for $a_{1}^{I}$ is 
\begin{align}
	H_{IJ}:=\frac{\partial^2 \Lambda^{(10-d)}}{\partial a_{1}^{I}\partial a_{1}^{J}}\sim \sum_{\rho}\rho^{I}\rho^{J}\cos \left[\pi (2n-1) \rho\cdot a_{1} \right]=\sum_{\rho}\rho^{I}\rho^{J}(-1)^{\rho\cdot a_{1}}.
\end{align}
The necessary condition for the critical point to be a minimum requires that all diagonal components of the Hessian be positive, i.e.,
\begin{align}\label{eq: condition for minimum}
\sum_{\rho\in\Delta_{+}}\left( \rho^{I}\right)^2 - \sum_{\rho\in\Delta_{-}}\left( \rho^{I}\right)^2 > 0~~~\text{for all $I=1,\cdots,8$}.
\end{align}
The question is whether a point $(E',a',a_{1})$ satisfying both eq.\eqref{eq: condition nF-nB=0} and eq.\eqref{eq: condition for minimum} exists or not. Actually, one can check that the examples given in the previous section that satisfy eq.\eqref{eq: condition nF-nB=0} do not satisfy eq.\eqref{eq: condition for minimum}. For $d=2$, since $|\Delta_{+}|=0$ and $|\Delta_{-}|=8$, this point corresponds to a maximum of the potential. For $d=3$, by eqs. \eqref{eq: d=3 n1 even odd}, we find
\begin{align}
	H_{11}>0,~~~H_{II}<0~~\text{for $I\neq 1$},
\end{align}
and the $a_{1}^{I\neq 1}$-directions are tachyonic. Thus, the point that we found in subsection \ref{subsec: suppression d=3} realizes the exponential suppression of the cosmological constant but corresponds to an unstable saddle point.

Also, in this paper, we just assume that the radion $R$, which has the runaway behavior, is sufficiently larger than the string length scale. 
It is an open question whether (meta)stable de Sitter vacua exist or not. In order to realize that, some extra ingredients (e.g. fluxes) are expected to be needed such as in refs. \cite{Basile:2018irz,Antonelli:2019nar,Basile:2021vxh}.

\section*{Acknowledgments}
The author would like to thank Yuta Hamada and Shun'ya Mizoguchi for valuable discussions, and Hiroshi Itoyama and Yuichi Koga for collaborations in related subjects. 
This work is supported in part by JSPS KAKENHI Grant Number 21J15497. 
\newpage
\appendix

\section{Lattices and characters}\label{app: lattice characters}

Irreducible representations of $SO(2n)$ ($D_{n}$) can be classified into the four conjugacy classes:
\begin{itemize}
	\item The trivial conjugacy class (the root lattice):
	\begin{align}
	\Gamma^{(n)}_{g}=\left\lbrace \left( n_1,\cdots,n_{n}\right) \left|~ n_{i}\in \mathbb{Z},~\sum_{i=1}^{n}n_{i}\in 2\mathbb{Z} \right. \right\rbrace .
	\end{align}
	\item The vector conjugacy class:
	\begin{align}
	\Gamma^{(n)}_{v}=\left\lbrace \left( n_1,\cdots,n_{n}\right) \left|~ n_{i}\in \mathbb{Z},~\sum_{i=1}^{n}n_{i}\in 2\mathbb{Z} +1\right. \right\rbrace .
	\end{align}
	\item The spinor conjugacy class:
	\begin{align}
	\Gamma^{(n)}_{s}=\left\lbrace \left( n_1+\frac{1}{2},\cdots,n_{n}+\frac{1}{2}\right) \left|~ n_{i}\in \mathbb{Z},~\sum_{i=1}^{n}n_{i}\in 2\mathbb{Z} \right. \right\rbrace .
	\end{align}
	\item The conjugate spinor conjugacy class:
	\begin{align}
	\Gamma^{(n)}_{c}=\left\lbrace \left( n_1+\frac{1}{2},\cdots,n_{n}+\frac{1}{2}\right) \left|~ n_{i}\in \mathbb{Z},~\sum_{i=1}^{n}n_{i}\in 2\mathbb{Z}+1 \right. \right\rbrace .
	\end{align}
\end{itemize}
The $E_{8}$ lattice is expressed in terms of the conjugacy classes of $SO(16)$:
\begin{align}
	\Gamma_{E_{8}}=\Gamma_{g}^{(8)}+\Gamma_{s}^{(8)}.
\end{align}
The $SO(2n)$ characters of the corresponding conjugacy classes are defined as
\begin{align}
O_{2n}
&=\frac{1}{\eta^{n}} \sum_{\pi\in \Gamma^{(n)}_{g}}q^{\frac{1}{2}|\pi|^2}
=\frac{1}{2\eta^{n}}\left( \vartheta_{3}^{n}(\tau)+ \vartheta_{4}^{n}(\tau)\right),\\
V_{2n}
&=\frac{1}{\eta^{n}} \sum_{\pi\in \Gamma^{(n)}_{v}}q^{\frac{1}{2}|\pi|^2}
=\frac{1}{2\eta^{n}}\left( \vartheta_{3}^{n}(\tau)- \vartheta_{4}^{n}(\tau)\right),\\
S_{2n}
&=\frac{1}{\eta^{n}} \sum_{\pi\in \Gamma^{(n)}_{s}}q^{\frac{1}{2}|\pi|^2}
=\frac{1}{2\eta^{n}}\left( \vartheta_{2}^{n}(\tau)+ \vartheta_{1}^{n}(\tau)
\right),\\
C_{2n}
&=\frac{1}{\eta^{n}} \sum_{\pi\in \Gamma^{(n)}_{c}}q^{\frac{1}{2}|\pi|^2}
=\frac{1}{2\eta^{n}}\left( \vartheta_{2}^{n}(\tau)-\vartheta_{1}^{n}(\tau)
\right),
\end{align}
where the Dedekind eta function and the theta functions with characteristics are defined as
\begin{align}
\eta(\tau)&=q^{\frac{1}{24}}\prod_{n=1}^{\infty}\left( 1-q^{n}\right),\\
\vartheta
\begin{bmatrix} 
\alpha\\ 
\beta\\ 
\end{bmatrix}(\tau)&=\sum_{n=-\infty}^{\infty}\exp\left( \pi i (n+\alpha)^2 \tau +2\pi i (n+\alpha)\beta\right),
\end{align} 
and
\begin{align}
	\vartheta_{3}=\vartheta
	\begin{bmatrix} 
	0\\ 
	0\\ 
	\end{bmatrix},~~~\vartheta_{4}=\vartheta
	\begin{bmatrix} 
	0\\
	\frac{1}{2}\\ 
	\end{bmatrix},~~~\vartheta_{2}=\vartheta
	\begin{bmatrix} 
	\frac{1}{2}\\ 
	0\\ 
	\end{bmatrix},~~~\vartheta_{1}=\vartheta
	\begin{bmatrix} 
	\frac{1}{2}\\ 
	\frac{1}{2}\\ 
	\end{bmatrix}.
\end{align}
The $SO(8)$ characters satisfy the Jacobi's abstruse identity
\begin{align}\label{eq: Jacobi identity}
V_{8}-S_{8}=0.
\end{align}

In order to check the modular invariance of the partition functions, it is useful to reveal how the $SO(2n)$ characters transform under $T:\tau\to\tau+1$ and $S:\tau\to-1/\tau$. The eta function and the theta function satisfy the following identities:
\begin{align}
\label{eq: T-trans of eta theta}
\eta (\tau+1)&=e^{\frac{\pi i}{12}}\eta(\tau),~~~~
\vartheta
\begin{bmatrix} 
\alpha\\ 
\beta\\ 
\end{bmatrix}(\tau+1)=e^{-\pi i\alpha(\alpha-1)}\vartheta
\begin{bmatrix} 
\alpha\\ 
\alpha+\beta-\frac{1}{2}\\ 
\end{bmatrix}(\tau),\\
\label{eq: S-trans of eta theta}
\eta \left(-\frac{1}{\tau} \right) &=(-i\tau)^{\frac{1}{2}}\eta(\tau),~~~~
\vartheta
\begin{bmatrix} 
\alpha\\ 
\beta\\ 
\end{bmatrix}\left(-\frac{1}{\tau} \right)=(-i\tau)^{\frac{1}{2}}e^{2\pi i\alpha\beta}\vartheta
\begin{bmatrix} 
\beta\\ 
-\alpha\\ 
\end{bmatrix}(\tau).
\end{align}
Then we find the transformation laws of the $SO(2n)$ characters:
\begin{align}
\label{eq: T-trans characters}
T:\left( O_{2n},V_{2n},S_{2n},C_{2n}\right) \to \left( O_{2n},V_{2n},S_{2n},C_{2n}\right) \mathcal{T}_{2n},\\
\label{eq: S-trans characters}
S:\left( O_{2n},V_{2n},S_{2n},C_{2n}\right) \to \left( O_{2n},V_{2n},S_{2n},C_{2n}\right) \mathcal{S}_{2n},
\end{align}
where $\mathcal{T}_{2n}$ and $\mathcal{S}_{2n}$ are $4\times 4$ matrices defined as follows:
\begin{align}
\mathcal{T}_{2n}=\begin{pmatrix}
e^{-\frac{n\pi i}{12}}&0&0&0\\
0&-e^{-\frac{n\pi i}{12}}&0&0\\
0&0&e^{\frac{n\pi i}{6}}&0\\
0&0&0&e^{\frac{n\pi i}{6}}\\
\end{pmatrix},~~~
\mathcal{S}_{2n}=\frac{1}{2}\begin{pmatrix}
1&1&1&1\\
1&1&-1&-1\\
1&-1&i^{n}&-i^{n}\\
1&-1&-i^{n}&i^{n}\\
\end{pmatrix}.
\end{align}

\section{Non-supersymmetric $E_8$ theory}
\label{app: non-susy E8}

In this appendix, we briefly review the non-supersymmetric $E_{8}$ theory constructed by the orbifold with $\R(-1)^F$. The starting point is the $10$D $E_{8}\times E'_{8}$ theory of which the partition function is given by 
\begin{align}
Z_{E_{8}\times E'_{8}}^{(10)}=Z_{B}^{(8)}\left( \bar{V}_{8}-\bar{S}_{8} \right) \sum_{\pi\in\Gamma_{E_{8}}}\sum_{\pi'\in\Gamma_{E'_{8}}}q^{\frac{1}{2}\left(\pi^2+\pi'^2 \right) }.
\end{align}
As in the CHL model, the partition function of the $E_{8}$ model is expressed as the sum of the building blocks:
\begin{align}
Z_{\mathcal{N}=0}^{(10)}=\frac{1}{2}Z_{B}^{(8)}\sum_{a,b=0,1}Z^{(10)}(g^a, g^b),
\end{align}
where
\begin{align}
Z^{(10)}(1,1)&=\left( \bar{V}_{8}-\bar{S}_{8} \right) \frac{1}{\eta^{16}}\sum_{\pi\in\Gamma_{E_{8}}}\sum_{\pi'\in\Gamma_{E'_{8}}}q^{\frac{1}{2}\left(\pi^2+\pi'^2 \right) },\\
Z^{(10)}(1,g)&=\left( \bar{V}_{8}+\bar{S}_{8} \right) \frac{1}{\eta^{16}}\left( \frac{2\eta^3}{\vartheta_{2}}\right)^{4} \sum_{\pi\in \sqrt{2}\Gamma_{E_{8}}}q^{\frac{1}{2}\pi^2 },\\
Z^{(10)}(g,1)&=\left( \bar{O}_{8}-\bar{C}_{8} \right) \frac{1}{\eta^{16}}\left( \frac{\eta^3}{\vartheta_{4}}\right)^{4} \sum_{\pi\in \frac{1}{\sqrt{2}}\Gamma_{E_{8}}} q^{\frac{1}{2}\pi^2 },\\
Z^{(10)}(g,g)&=\left( \bar{O}_{8}+\bar{C}_{8} \right) \frac{1}{\eta^{16}}\left( \frac{\eta^3}{\vartheta_{3}}\right)^{4} 
\sum_{\pi\in \frac{1}{\sqrt{2}}\Gamma_{E_{8}}}q^{\frac{1}{2}\pi^2 }e^{\pi i \pi^2}.
\end{align}
The total partition function is then written as
\begin{align}
\label{eq: partition function of E8}
Z_{\mathcal{N}=0}^{(10)}=\frac{1}{2}Z_{B}^{(8)}\frac{1}{\eta^{16}}&\left\lbrace 
\bar{V}_{8}\left(\sum_{\pi\in\Gamma_{E_{8}}}\sum_{\pi'\in\Gamma_{E'_{8}}}q^{\frac{1}{2}\left(\pi^2+\pi'^2 \right) }  +\left( \frac{2\eta^3}{\vartheta_{2}}\right)^{4} \sum_{\pi\in \sqrt{2}\Gamma_{E_{8}}} q^{\frac{1}{2}\pi^2 }\right) \right. \nonumber\\
&\left. 
-\bar{S}_{8}\left(\sum_{\pi\in\Gamma_{E_{8}}}\sum_{\pi'\in\Gamma_{E'_{8}}}q^{\frac{1}{2}\left(\pi^2+\pi'^2 \right) }  -\left( \frac{2\eta^3}{\vartheta_{2}}\right)^{4} \sum_{\pi\in \sqrt{2}\Gamma_{E_{8}}} q^{\frac{1}{2}\pi^2 }\right) \right. \nonumber\\
&\left. 
+\bar{O}_{8} \sum_{\pi\in \frac{1}{\sqrt{2}}\Gamma_{E_{8}}} q^{\frac{1}{2}\pi^2 }
\left(\left( \frac{\eta^3}{\vartheta_{4}}\right)^{4}+\left( \frac{\eta^3}{\vartheta_{3}}\right)^{4}e^{\pi i \pi^2}\right) \right. \nonumber\\
&\left. 
-\bar{C}_{8} \sum_{\pi\in \frac{1}{\sqrt{2}}\Gamma_{E_{8}}} q^{\frac{1}{2}\pi^2 }
\left(\left( \frac{\eta^3}{\vartheta_{4}}\right)^{4}-\left( \frac{\eta^3}{\vartheta_{3}}\right)^{4}e^{\pi i \pi^2}\right)
\right\rbrace.
\end{align}
The spectrum can be identified in the same way as in the CHL model. In the massless level, there are gauge bosons of $E_8$ and non-chiral fermions transforming in the adjoint representation of the $E_{8}$. The tachyonic state, which is a singlet of the $E_{8}$, also exists. The spectrum of the non-supersymmetric $E_8$ string is summarized in table \ref{table: spectrum of E8 string}.
\begin{table}[t]
	\centering
	\begin{tabular}{c||c|c|c|c}
		\hline
		spacetime $SO(8)$ rep&$\boldsymbol{8}_{V}$ & $\boldsymbol{8}_{S}$  & $\boldsymbol{8}_{C}$  & $\boldsymbol{1}$   \\
		$E_{8}$ rep.&$\boldsymbol{248}$ & $\boldsymbol{248}$ & $\boldsymbol{248}$ & $\boldsymbol{1}$\\
		$M_{R}^2=M_{L}^2$&$0$ & $0$ & $0$ & $-1$\\
		\hline
	\end{tabular}
	\caption{The spectrum of the $E_{8}$ theory. The vector, spinor and co-spinor are massless while the scalar is tachyonic.}
		\label{table: spectrum of E8 string}
\end{table}

\section{Solutions in $d=2$}\label{app: table d=2}
In this appendix, we present the tables which summarize the solutions of the massless conditions in the case $d=2$ studied in subsection \ref{subsec: d=2}.

\subsection{Enhancement to $C_{9}\times C_{1}$}
The solutions in the scalar conjugacy class are those of eq.\eqref{eq: n integer ex1 d=2} and eq.\eqref{eq: condition short roots ex1 d=2} with $w^{1,2}\in\mathbb{Z}$ and $\pi_{+}\in\Gamma_{E_{8}}$ or $(w^1,w^2;\pi_{+})=(0,0;0)$ shifted by the following four vectors $v_a$:
\begin{align}
	v_1&=-v_2=\left( \frac{3}{2},1; -3,2; \frac{1}{2},-\frac{1}{2},-\frac{1}{2},-\frac{1}{2},-\frac{1}{2},-\frac{1}{2},-\frac{1}{2},\frac{5}{2}\right),\\
	v_3&=-v_4=\left( \frac{3}{2},2; -3,1; \frac{1}{2},-\frac{1}{2},-\frac{1}{2},-\frac{1}{2},-\frac{1}{2},-\frac{1}{2},-\frac{1}{2},\frac{5}{2}\right).
\end{align}
\begin{center}
	\begin{longtable}{|l|l|l|}
		\caption{The solutions of the massless conditions with the moduli \eqref{eq: moduli1 d=2}. } \label{table: C9 x C1 weights} \\
		
		\hline \multicolumn{1}{|c|}{\textbf{$\bm{SO(8)}$ rep.}} & \multicolumn{1}{c|}{\textbf{$\bm{P_{L}^2}$}} & \multicolumn{1}{c|}{\textbf{$\bm{\left(w;n;\pi_{+}\right)}$}} \\ \hline 
		\endfirsthead
		
		\multicolumn{3}{c}%
		{{\bfseries \tablename\ \thetable{} -- continued from previous page}} \\
		\hline \multicolumn{1}{|c|}{\textbf{$\bm{SO(8)}$ rep.}} & \multicolumn{1}{c|}{$\bm{P_{L}^2}$} & \multicolumn{1}{c|}{$\bm{\left(w;n;\pi_{+}\right)}$} \\ \hline 
		\endhead
		
		\hline \multicolumn{3}{|r|}{{Continued on next page}} \\ \hline
		\endfoot
		
		\hline \hline
		\endlastfoot
		
		$\boldsymbol{8}_{V}$, $\boldsymbol{8}_{S}$&$1$&$\pm\left( 0,0; 0,0; 1, \underline{1,0,0,0,0,0},0\right)$\\
		$\boldsymbol{8}_{V}$, $\boldsymbol{8}_{S}$&$1$&$\pm\left( 0,0; -1,0; 1, 0,0,0,0,0,0,1\right)$\\
		$\boldsymbol{8}_{V}$, $\boldsymbol{8}_{S}$&$1$&$\left( 0,0; 0,0; 0, \underline{1,-1,0,0,0,0},0\right)$\\
		$\boldsymbol{8}_{V}$, $\boldsymbol{8}_{S}$&$1$&$\left( 0,0; 1,0; 0, \underline{1,0,0,0,0,0},-1\right)$\\
		$\boldsymbol{8}_{V}$, $\boldsymbol{8}_{S}$&$1$&$\pm\left( 0,0; 0,0;  -\frac{1}{2},\underline{\frac{1}{2},\frac{1}{2},\frac{1}{2},\frac{1}{2},-\frac{1}{2}},\frac{1}{2}\right) $ \\ 
		$\boldsymbol{8}_{V}$, $\boldsymbol{8}_{S}$&$1$&$\pm\left( 0,0; 1,0;  -\frac{1}{2},\frac{1}{2},\frac{1}{2},\frac{1}{2},\frac{1}{2},\frac{1}{2},-\frac{1}{2}\right) $ \\ 
		$\boldsymbol{8}_{V}$, $\boldsymbol{8}_{S}$&$1$&$\pm\left( 0,0; 0,0;  \frac{1}{2},\frac{1}{2},\frac{1}{2},\frac{1}{2},\frac{1}{2},\frac{1}{2},\frac{1}{2}\right) $ \\ 
		$\boldsymbol{8}_{V}$, $\boldsymbol{8}_{S}$&$1$&$\pm\left( 1,1; -2,1;  0,\underline{-1,-1,0,0,0,0},2\right) $ \\ 
		$\boldsymbol{8}_{V}$, $\boldsymbol{8}_{S}$&$1$&$\pm\left( 1,1; -2,1;  1,\underline{-1,0,0,0,0,0},2\right) $ \\ 
		$\boldsymbol{8}_{V}$, $\boldsymbol{8}_{S}$&$1$&$\pm\left( 1,1; -1,1;  0,\underline{-1,0,0,0,0,0},1\right) $ \\
		$\boldsymbol{8}_{V}$, $\boldsymbol{8}_{S}$&$1$&$\pm\left( 1,1; -1,1;  1,0,0,0,0,0,0,1\right) $ \\  
		$\boldsymbol{8}_{V}$, $\boldsymbol{8}_{S}$&$1$&$\pm\left( 1,1; -1,1;  \frac{1}{2},-\frac{1}{2},-\frac{1}{2},-\frac{1}{2},-\frac{1}{2},-\frac{1}{2},-\frac{1}{2},\frac{1}{2}\right) $ \\  
		$\boldsymbol{8}_{V}$, $\boldsymbol{8}_{S}$&$1$&$\pm\left( 1,1; -2,1;  \frac{3}{2},-\frac{1}{2},-\frac{1}{2},-\frac{1}{2},-\frac{1}{2},-\frac{1}{2},-\frac{1}{2},\frac{3}{2}\right) $ \\  
		$\boldsymbol{8}_{V}$, $\boldsymbol{8}_{S}$&$1$&$\pm\left( 1,1; -2,1;  \frac{1}{2},\underline{-\frac{3}{2},-\frac{1}{2},-\frac{1}{2},-\frac{1}{2},-\frac{1}{2},-\frac{1}{2}},\frac{3}{2}\right) $ \\  
		$\boldsymbol{8}_{V}$&$2$&$\pm\left( 0,1; -2,1;  0,0,0,0,0,0,0,0\right) $ \\ 
		$\boldsymbol{8}_{V}$&$2$&$\pm\left( 1,1; 0,1;  0,0,0,0,0,0,0,0\right) $ \\ 
		$\boldsymbol{8}_{V}$&$2$&$\pm\left( 1,1; -2,1;  0,\underline{-2,0,0,0,0,0},2\right) $ \\ 
		$\boldsymbol{8}_{V}$&$2$&$\pm\left( 1,1; -2,1;  2,0,0,0,0,0,0,2\right) $ \\ 
		$\boldsymbol{8}_{V}$&$2$&$\pm\left( 1,1; -2,1;  1,-1,-1,-1,-1,-1,-1,1\right)$ \\ 
		$\boldsymbol{8}_{C}$&$1$&$\pm\left( \frac{1}{2},0; 1,0; 0,0,0,0,0,0,0,0\right)$ \\
		$\boldsymbol{8}_{C}$&$1$&$\pm\left( \frac{1}{2},0; 0,0; 0,\underline{-1,0,0,0,0,0},1\right)$ \\
		$\boldsymbol{8}_{C}$&$1$&$\pm\left( \frac{1}{2},0; 0,0; 1,0,0,0,0,0,0,1\right)$ \\
		$\boldsymbol{8}_{C}$&$1$&$\pm\left( \frac{1}{2},0; 0,0; \frac{1}{2},-\frac{1}{2},-\frac{1}{2},-\frac{1}{2},-\frac{1}{2},-\frac{1}{2},-\frac{1}{2},\frac{1}{2}\right)$ \\
		$\boldsymbol{8}_{C}$&$1$&$\pm\left( \frac{1}{2},1; -1,1; 0,0,0,0,0,0,0,0\right)$ \\
		$\boldsymbol{8}_{C}$&$1$&$\pm\left( \frac{1}{2},1; -2,1; 0,\underline{-1,0,0,0,0,0},1\right)$ \\
		$\boldsymbol{8}_{C}$&$1$&$\pm\left( \frac{1}{2},1; -2,1; 1,0,0,0,0,0,0,1\right)$ \\
		$\boldsymbol{8}_{C}$&$1$&$\pm\left( \frac{1}{2},1; -2,1; \frac{1}{2},-\frac{1}{2},-\frac{1}{2},-\frac{1}{2},-\frac{1}{2},-\frac{1}{2},-\frac{1}{2},\frac{1}{2}\right)$ \\
		$\boldsymbol{1}$&$0$&$ v_{1,2}=\pm\left( \frac{3}{2},1; -3,2; \frac{1}{2},-\frac{1}{2},-\frac{1}{2},-\frac{1}{2},-\frac{1}{2},-\frac{1}{2},-\frac{1}{2},\frac{5}{2}\right)$ \\
		$\boldsymbol{1}$&$0$&$v_{3,2}=\pm\left( \frac{3}{2},2; -3,1; \frac{1}{2},-\frac{1}{2},-\frac{1}{2},-\frac{1}{2},-\frac{1}{2},-\frac{1}{2},-\frac{1}{2},\frac{5}{2}\right)$ \\
		$\boldsymbol{1}$&$1$&$\pm\left( 0,0; 0,0; 1, \underline{1,0,0,0,0,0},0\right)+v_{a}$\\
		$\boldsymbol{1}$&$1$&$\pm\left( 0,0; -1,0; 1, 0,0,0,0,0,0,1\right)+v_{a}$\\
		$\boldsymbol{1}$&$1$&$\left( 0,0; 0,0; 0, \underline{1,-1,0,0,0,0},0\right)+v_{a}$\\
		$\boldsymbol{1}$&$1$&$\left( 0,0; 1,0; 0, \underline{1,0,0,0,0,0},-1\right)+v_{a}$\\
		$\boldsymbol{1}$&$1$&$\pm\left( 0,0; 0,0;  -\frac{1}{2},\underline{\frac{1}{2},\frac{1}{2},\frac{1}{2},\frac{1}{2},-\frac{1}{2}},\frac{1}{2}\right)+v_{a} $ \\ 
		$\boldsymbol{1}$&$1$&$\pm\left( 0,0; 1,0;  -\frac{1}{2},\frac{1}{2},\frac{1}{2},\frac{1}{2},\frac{1}{2},\frac{1}{2},-\frac{1}{2}\right) +v_{a}$ \\ 
		$\boldsymbol{1}$&$1$&$\pm\left( 0,0; 0,0;  \frac{1}{2},\frac{1}{2},\frac{1}{2},\frac{1}{2},\frac{1}{2},\frac{1}{2},\frac{1}{2}\right) +v_{a}$ \\ 
		$\boldsymbol{1}$&$1$&$\pm\left( 1,1; -2,1;  0,\underline{-1,-1,0,0,0,0},2\right) +v_{a}$ \\ 
		$\boldsymbol{1}$&$1$&$\pm\left( 1,1; -2,1;  1,\underline{-1,0,0,0,0,0},2\right) +v_{a}$ \\ 
		$\boldsymbol{1}$&$1$&$\pm\left( 1,1; -1,1;  0,\underline{-1,0,0,0,0,0},1\right) +v_{a}$ \\
		$\boldsymbol{1}$&$1$&$\pm\left( 1,1; -1,1;  1,0,0,0,0,0,0,1\right) +v_{a}$ \\  
		$\boldsymbol{1}$&$1$&$\pm\left( 1,1; -1,1;  \frac{1}{2},-\frac{1}{2},-\frac{1}{2},-\frac{1}{2},-\frac{1}{2},-\frac{1}{2},-\frac{1}{2},\frac{1}{2}\right) +v_{a}$ \\  
		$\boldsymbol{1}$&$1$&$\pm\left( 1,1; -2,1;  \frac{3}{2},-\frac{1}{2},-\frac{1}{2},-\frac{1}{2},-\frac{1}{2},-\frac{1}{2},-\frac{1}{2},\frac{3}{2}\right) +v_{a}$ \\  
		$\boldsymbol{1}$&$1$&$\pm\left( 1,1; -2,1;  \frac{1}{2},\underline{-\frac{3}{2},-\frac{1}{2},-\frac{1}{2},-\frac{1}{2},-\frac{1}{2},-\frac{1}{2}},\frac{3}{2}\right) +v_{a}$ \\  
		\hline
	\end{longtable}
\end{center}

\subsection{Enhancement to $A_{4}\times C_{4}\times C_{2}$}
In this example, there are no solutions in the scalar conjugacy class.
\begin{center}
	\begin{longtable}{|l|l|l|}
		\caption{The solutions of the massless conditions with the moduli \eqref{eq: moduli2 d=2}. } \label{table: A4 x C4 x C2 weights} \\
		
		\hline \multicolumn{1}{|c|}{\textbf{$\bm{SO(8)}$ rep.}} & \multicolumn{1}{c|}{\textbf{$\bm{P_{L}^2}$}} & \multicolumn{1}{c|}{\textbf{$\bm{\left(w;n;\pi_{+}\right)}$}} \\ \hline 
		\endfirsthead
		
		\multicolumn{3}{c}%
		{{\bfseries \tablename\ \thetable{} -- continued from previous page}} \\
		\hline \multicolumn{1}{|c|}{\textbf{$\bm{SO(8)}$ rep.}} & \multicolumn{1}{c|}{$\bm{P_{L}^2}$} & \multicolumn{1}{c|}{$\bm{\left(w;n;\pi_{+}\right)}$} \\ \hline 
		\endhead
		
		\hline \multicolumn{3}{|r|}{{Continued on next page}} \\ \hline
		\endfoot
		
		\hline \hline
		\endlastfoot
		
		$\boldsymbol{8}_{V}$, $\boldsymbol{8}_{S}$&$1$&$\left( 0,0; 0,0; \underline{\pm1,\pm1,0},0,0,0,0,0\right)$\\
		$\boldsymbol{8}_{V}$, $\boldsymbol{8}_{S}$&$1$&$\pm\left( 0,0; 0,0;  \underline{\pm\frac{1}{2},\pm\frac{1}{2},\pm\frac{1}{2}}_{+},-\frac{1}{2},-\frac{1}{2},-\frac{1}{2},-\frac{1}{2},-\frac{1}{2}\right) $ \\ 
		$\boldsymbol{8}_{V}$, $\boldsymbol{8}_{S}$&$1$&$\left( 0,0; 0,0; 0,0,0,\underline{+1,-1,0},0,0\right)$ \\
		$\boldsymbol{8}_{V}$, $\boldsymbol{8}_{S}$&$1$&$\pm\left( 0,1; -1,-1; 0,0,0,\underline{-1,-1,0},0,2\right)$ \\
		$\boldsymbol{8}_{V}$, $\boldsymbol{8}_{S}$&$1$&$\pm\left( 1,2;-1,-2; 0,0,0,\underline{-1,-1,0},-2,4\right)$ \\
		$\boldsymbol{8}_{V}$, $\boldsymbol{8}_{S}$&$1$&$\pm\left( 1,3; -2,-3; 0,0,0,\underline{-1,-1,-2},-2,6\right)$ \\
		$\boldsymbol{8}_{V}$, $\boldsymbol{8}_{S}$&$1$&$\pm\left( 0,0; 1,1; 0,0,0,0,0,0,1,-1\right)$ \\
		$\boldsymbol{8}_{V}$, $\boldsymbol{8}_{S}$&$1$&$\pm\left( 0,1; -1,0; 0,0,0,0,0,0,-1,1\right)$ \\
		$\boldsymbol{8}_{V}$&$2$&$\pm\left( 0,1;0,1; 0,0,0,0,0,0,0,0\right)$ \\
		$\boldsymbol{8}_{V}$&$2$&$\pm\left( 0,1; -2,-1; 0,0,0,0,0,0,-2,2\right)$ \\
		$\boldsymbol{8}_{V}$&$2$&$\pm\left( 1,1; 0,-1; 0,0,0,0,0,0,-2,2\right)$ \\
		$\boldsymbol{8}_{V}$&$2$&$\pm\left( 1,3;-2,-3;0,0,0,\underline{-2,-2,0},-2,6\right)$ \\
		$\boldsymbol{8}_{S}$&$2$&$\pm\left( 0,1;-1,-1;0,0,0,\underline{-2,0,0},0,2\right)$ \\
		$\boldsymbol{8}_{S}$&$2$&$\pm\left( 1,2;-1,-2;0,0,0,\underline{-2,0,0},-2,4\right)$ \\
		$\boldsymbol{8}_{S}$&$2$&$\pm\left( 1,4;-3,-4;0,0,0,-2,-2,-2,-2,8\right)$ \\
		$\boldsymbol{8}_{S}$&$2$&$\pm\left( 2,5;-3,-5;0,0,0,-2,-2,-2,-4,10\right)$ \\
		$\boldsymbol{8}_{C}$&$1$&$\pm\left( \frac{1}{2},1; -1,-2; 0,0,0,\underline{-1,-1,0},-1,3\right)$\\
		$\boldsymbol{8}_{C}$&$1$&$\pm\left( \frac{1}{2},1; 0,-1; 0,0,0,\underline{-1,-1,0},0,2\right)$\\
		$\boldsymbol{8}_{C}$&$1$&$\pm\left( \frac{1}{2},1; 0,0; 0,0,0,0,0,0,-1,1\right)$\\
		$\boldsymbol{8}_{C}$&$1$&$\pm\left( \frac{1}{2},1; -1,-1; 0,0,0,0,0,0,-2,2\right)$\\
		$\boldsymbol{8}_{C}$&$1$&$\pm\left( \frac{1}{2},2; -2,-2; 0,0,0,\underline{-1,-1,0},-2,4\right)$\\
		$\boldsymbol{8}_{C}$&$1$&$\pm\left( \frac{1}{2},2; -1,-1; 0,0,0,\underline{-1,-1,0},-1,3\right)$\\
		$\boldsymbol{8}_{C}$&$1$&$\pm\left( \frac{1}{2},0; 1,0; 0,0,0,0,0,0,0,0\right)$\\
		$\boldsymbol{8}_{C}$&$1$&$\pm\left( \frac{1}{2},0; 1,0; 0,0,0,0,0,0,-1,1\right)$\\\hline
	\end{longtable}
\end{center}


\clearpage
\bibliographystyle{elsarticle-num}
\bibliography{ref}

\begin{thebibliography}{10}
\expandafter\ifx\csname url\endcsname\relax
  \def\url#1{\texttt{#1}}\fi
\expandafter\ifx\csname urlprefix\endcsname\relax\def\urlprefix{URL }\fi
\expandafter\ifx\csname href\endcsname\relax
  \def\href#1#2{#2} \def\path#1{#1}\fi

\bibitem{Gross:1984dd}
D.~J. Gross, J.~A. Harvey, E.~J. Martinec, R.~Rohm, {The Heterotic String},
  Phys. Rev. Lett. 54 (1985) 502--505.
\newblock \href {https://doi.org/10.1103/PhysRevLett.54.502}
  {\path{doi:10.1103/PhysRevLett.54.502}}.

\bibitem{Dixon:1986iz}
L.~J. Dixon, J.~A. Harvey, {String Theories in Ten-Dimensions Without
  Space-Time Supersymmetry}, Nucl. Phys. B 274 (1986) 93--105.
\newblock \href {https://doi.org/10.1016/0550-3213(86)90619-X}
  {\path{doi:10.1016/0550-3213(86)90619-X}}.

\bibitem{AlvarezGaume:1986jb}
L.~Alvarez-Gaume, P.~H. Ginsparg, G.~W. Moore, C.~Vafa, {An O(16) x O(16)
  Heterotic String}, Phys. Lett. B 171 (1986) 155--162.
\newblock \href {https://doi.org/10.1016/0370-2693(86)91524-8}
  {\path{doi:10.1016/0370-2693(86)91524-8}}.

\bibitem{Ginsparg:1986wr}
P.~H. Ginsparg, C.~Vafa, {Toroidal Compactification of Nonsupersymmetric
  Heterotic Strings}, Nucl. Phys. B 289 (1987) 414.
\newblock \href {https://doi.org/10.1016/0550-3213(87)90387-7}
  {\path{doi:10.1016/0550-3213(87)90387-7}}.

\bibitem{Vafa:1986wx}
C.~Vafa, {Modular Invariance and Discrete Torsion on Orbifolds}, Nucl. Phys. B
  273 (1986) 592--606.
\newblock \href {https://doi.org/10.1016/0550-3213(86)90379-2}
  {\path{doi:10.1016/0550-3213(86)90379-2}}.

\bibitem{Blaszczyk:2014qoa}
M.~Blaszczyk, S.~Groot~Nibbelink, O.~Loukas, S.~Ramos-Sanchez,
  {Non-supersymmetric heterotic model building}, JHEP 10 (2014) 119.
\newblock \href {http://arxiv.org/abs/1407.6362} {\path{arXiv:1407.6362}},
  \href {https://doi.org/10.1007/JHEP10(2014)119}
  {\path{doi:10.1007/JHEP10(2014)119}}.

\bibitem{Hamada:2015ria}
Y.~Hamada, H.~Kawai, K.-y. Oda, {Eternal Higgs inflation and the cosmological
  constant problem}, Phys. Rev. D 92 (2015) 045009.
\newblock \href {http://arxiv.org/abs/1501.04455} {\path{arXiv:1501.04455}},
  \href {https://doi.org/10.1103/PhysRevD.92.045009}
  {\path{doi:10.1103/PhysRevD.92.045009}}.

\bibitem{Perez-Martinez:2021zjj}
R.~Perez-Martinez, S.~Ramos-Sanchez, P.~K.~S. Vaudrevange, {Landscape of
  promising nonsupersymmetric string models}, Phys. Rev. D 104~(4) (2021)
  046026.
\newblock \href {http://arxiv.org/abs/2105.03460} {\path{arXiv:2105.03460}},
  \href {https://doi.org/10.1103/PhysRevD.104.046026}
  {\path{doi:10.1103/PhysRevD.104.046026}}.

\bibitem{Baykara:2022cwj}
Z.~K. Baykara, D.~Robbins, S.~Sethi, {Non-Supersymmetric AdS from String
  Theory} (12 2022).
\newblock \href {http://arxiv.org/abs/2212.02557} {\path{arXiv:2212.02557}}.

\bibitem{Ashfaque:2015vta}
J.~M. Ashfaque, P.~Athanasopoulos, A.~E. Faraggi, H.~Sonmez, {Non-Tachyonic
  Semi-Realistic Non-Supersymmetric Heterotic String Vacua}, Eur. Phys. J. C
  76~(4) (2016) 208.
\newblock \href {http://arxiv.org/abs/1506.03114} {\path{arXiv:1506.03114}},
  \href {https://doi.org/10.1140/epjc/s10052-016-4056-2}
  {\path{doi:10.1140/epjc/s10052-016-4056-2}}.

\bibitem{Faraggi:2019drl}
A.~E. Faraggi, V.~G. Matyas, B.~Percival, {Stable Three Generation
  Standard--like Model From a Tachyonic Ten Dimensional Heterotic--String
  Vacuum}, Eur. Phys. J. C 80~(4) (2020) 337.
\newblock \href {http://arxiv.org/abs/1912.00061} {\path{arXiv:1912.00061}},
  \href {https://doi.org/10.1140/epjc/s10052-020-7894-x}
  {\path{doi:10.1140/epjc/s10052-020-7894-x}}.

\bibitem{Faraggi:2019fap}
A.~E. Faraggi, {String Phenomenology From a Worldsheet Perspective}, Eur. Phys.
  J. C 79~(8) (2019) 703.
\newblock \href {http://arxiv.org/abs/1906.09448} {\path{arXiv:1906.09448}},
  \href {https://doi.org/10.1140/epjc/s10052-019-7222-5}
  {\path{doi:10.1140/epjc/s10052-019-7222-5}}.

\bibitem{Faraggi:2020hpy}
A.~E. Faraggi, V.~G. Matyas, B.~Percival, {Type $\mathbf{\bar{0}}$ heterotic
  string orbifolds}, Phys. Lett. B 814 (2021) 136080.
\newblock \href {http://arxiv.org/abs/2011.12630} {\path{arXiv:2011.12630}},
  \href {https://doi.org/10.1016/j.physletb.2021.136080}
  {\path{doi:10.1016/j.physletb.2021.136080}}.

\bibitem{Faraggi:2020wej}
A.~E. Faraggi, V.~G. Matyas, B.~Percival, {Towards the Classification of
  Tachyon-Free Models From Tachyonic Ten-Dimensional Heterotic String Vacua},
  Nucl. Phys. B 961 (2020) 115231.
\newblock \href {http://arxiv.org/abs/2006.11340} {\path{arXiv:2006.11340}},
  \href {https://doi.org/10.1016/j.nuclphysb.2020.115231}
  {\path{doi:10.1016/j.nuclphysb.2020.115231}}.

\bibitem{Faraggi:2020wld}
A.~E. Faraggi, V.~G. Matyas, B.~Percival, {Classification of nonsupersymmetric
  Pati-Salam heterotic string models}, Phys. Rev. D 104~(4) (2021) 046002.
\newblock \href {http://arxiv.org/abs/2011.04113} {\path{arXiv:2011.04113}},
  \href {https://doi.org/10.1103/PhysRevD.104.046002}
  {\path{doi:10.1103/PhysRevD.104.046002}}.

\bibitem{Faraggi:2021mws}
A.~E. Faraggi, B.~Percival, S.~Schewe, D.~Wojtczak, {Satisfiability modulo
  theories and chiral heterotic string vacua with positive cosmological
  constant}, Phys. Lett. B 816 (2021) 136187.
\newblock \href {http://arxiv.org/abs/2101.03227} {\path{arXiv:2101.03227}},
  \href {https://doi.org/10.1016/j.physletb.2021.136187}
  {\path{doi:10.1016/j.physletb.2021.136187}}.

\bibitem{Itoyama:1986ei}
H.~Itoyama, T.~R. Taylor, {Supersymmetry Restoration in the Compactified O(16)
  x O(16)-prime Heterotic String Theory}, Phys. Lett. B 186 (1987) 129--133.
\newblock \href {https://doi.org/10.1016/0370-2693(87)90267-X}
  {\path{doi:10.1016/0370-2693(87)90267-X}}.

\bibitem{Blum:1997gw}
J.~D. Blum, K.~R. Dienes, {Strong / weak coupling duality relations for
  nonsupersymmetric string theories}, Nucl. Phys. B 516 (1998) 83--159.
\newblock \href {http://arxiv.org/abs/hep-th/9707160}
  {\path{arXiv:hep-th/9707160}}, \href
  {https://doi.org/10.1016/S0550-3213(97)00803-1}
  {\path{doi:10.1016/S0550-3213(97)00803-1}}.

\bibitem{Scherk:1978ta}
J.~Scherk, J.~H. Schwarz, {Spontaneous Breaking of Supersymmetry Through
  Dimensional Reduction}, Phys. Lett. B 82 (1979) 60--64.
\newblock \href {https://doi.org/10.1016/0370-2693(79)90425-8}
  {\path{doi:10.1016/0370-2693(79)90425-8}}.

\bibitem{Rohm:1983aq}
R.~Rohm, {Spontaneous Supersymmetry Breaking in Supersymmetric String
  Theories}, Nucl. Phys. B 237 (1984) 553--572.
\newblock \href {https://doi.org/10.1016/0550-3213(84)90007-5}
  {\path{doi:10.1016/0550-3213(84)90007-5}}.

\bibitem{Kounnas:1989dk}
C.~Kounnas, B.~Rostand, {Coordinate Dependent Compactifications and Discrete
  Symmetries}, Nucl. Phys. B 341 (1990) 641--665.
\newblock \href {https://doi.org/10.1016/0550-3213(90)90543-M}
  {\path{doi:10.1016/0550-3213(90)90543-M}}.

\bibitem{Faraggi:2009xy}
A.~E. Faraggi, M.~Tsulaia, {Interpolations Among NAHE-based Supersymmetric and
  Nonsupersymmetric String Vacua}, Phys. Lett. B 683 (2010) 314--320.
\newblock \href {http://arxiv.org/abs/0911.5125} {\path{arXiv:0911.5125}},
  \href {https://doi.org/10.1016/j.physletb.2009.12.039}
  {\path{doi:10.1016/j.physletb.2009.12.039}}.

\bibitem{Florakis:2021bws}
I.~Florakis, J.~Rizos, K.~Violaris-Gountonis, {Super no-scale models with
  Pati-Salam gauge group}, Nucl. Phys. B 976 (2022) 115689.
\newblock \href {http://arxiv.org/abs/2110.06752} {\path{arXiv:2110.06752}},
  \href {https://doi.org/10.1016/j.nuclphysb.2022.115689}
  {\path{doi:10.1016/j.nuclphysb.2022.115689}}.

\bibitem{Florakis:2022avh}
I.~Florakis, J.~Rizos, K.~Violaris-Gountonis, {Three-generation super no-scale
  models in heterotic superstrings}, Phys. Lett. B 833 (2022) 137311.
\newblock \href {http://arxiv.org/abs/2206.09732} {\path{arXiv:2206.09732}},
  \href {https://doi.org/10.1016/j.physletb.2022.137311}
  {\path{doi:10.1016/j.physletb.2022.137311}}.

\bibitem{Abel:2015oxa}
S.~Abel, K.~R. Dienes, E.~Mavroudi, {Towards a nonsupersymmetric string
  phenomenology}, Phys. Rev. D 91~(12) (2015) 126014.
\newblock \href {http://arxiv.org/abs/1502.03087} {\path{arXiv:1502.03087}},
  \href {https://doi.org/10.1103/PhysRevD.91.126014}
  {\path{doi:10.1103/PhysRevD.91.126014}}.

\bibitem{Aaronson:2016kjm}
B.~Aaronson, S.~Abel, E.~Mavroudi, {Interpolations from supersymmetric to
  nonsupersymmetric strings and their properties}, Phys. Rev. D 95~(10) (2017)
  106001.
\newblock \href {http://arxiv.org/abs/1612.05742} {\path{arXiv:1612.05742}},
  \href {https://doi.org/10.1103/PhysRevD.95.106001}
  {\path{doi:10.1103/PhysRevD.95.106001}}.

\bibitem{Abel:2017vos}
S.~Abel, K.~R. Dienes, E.~Mavroudi, {GUT precursors and entwined SUSY: The
  phenomenology of stable nonsupersymmetric strings}, Phys. Rev. D 97~(12)
  (2018) 126017.
\newblock \href {http://arxiv.org/abs/1712.06894} {\path{arXiv:1712.06894}},
  \href {https://doi.org/10.1103/PhysRevD.97.126017}
  {\path{doi:10.1103/PhysRevD.97.126017}}.

\bibitem{Itoyama:1987rc}
H.~Itoyama, T.~R. Taylor, {Small Cosmological Constant in String Models}, in:
  {International Europhysics Conference on High-energy Physics}, 1987.

\bibitem{Kounnas:2016gmz}
C.~Kounnas, H.~Partouche, {Super no-scale models in string theory}, Nucl. Phys.
  B 913 (2016) 593--626.
\newblock \href {http://arxiv.org/abs/1607.01767} {\path{arXiv:1607.01767}},
  \href {https://doi.org/10.1016/j.nuclphysb.2016.10.001}
  {\path{doi:10.1016/j.nuclphysb.2016.10.001}}.

\bibitem{Kounnas:2017mad}
C.~Kounnas, H.~Partouche, {$\mathcal N=2 \to 0$ super no-scale models and
  moduli quantum stability}, Nucl. Phys. B 919 (2017) 41--73.
\newblock \href {http://arxiv.org/abs/1701.00545} {\path{arXiv:1701.00545}},
  \href {https://doi.org/10.1016/j.nuclphysb.2017.03.011}
  {\path{doi:10.1016/j.nuclphysb.2017.03.011}}.

\bibitem{Coudarchet:2017pie}
T.~Coudarchet, C.~Fleming, H.~Partouche, {Quantum no-scale regimes in string
  theory}, Nucl. Phys. B 930 (2018) 235--254.
\newblock \href {http://arxiv.org/abs/1711.09122} {\path{arXiv:1711.09122}},
  \href {https://doi.org/10.1016/j.nuclphysb.2018.03.002}
  {\path{doi:10.1016/j.nuclphysb.2018.03.002}}.

\bibitem{Coudarchet:2018ztz}
T.~Coudarchet, H.~Partouche, {Quantum no-scale regimes and moduli dynamics},
  Nucl. Phys. B 933 (2018) 134--184.
\newblock \href {http://arxiv.org/abs/1804.00466} {\path{arXiv:1804.00466}},
  \href {https://doi.org/10.1016/j.nuclphysb.2018.06.009}
  {\path{doi:10.1016/j.nuclphysb.2018.06.009}}.

\bibitem{Itoyama:2019yst}
H.~Itoyama, S.~Nakajima, {Exponentially suppressed cosmological constant with
  enhanced gauge symmetry in heterotic interpolating models}, PTEP 2019~(12)
  (2019) 123B01.
\newblock \href {http://arxiv.org/abs/1905.10745} {\path{arXiv:1905.10745}},
  \href {https://doi.org/10.1093/ptep/ptz123} {\path{doi:10.1093/ptep/ptz123}}.

\bibitem{Itoyama:2020ifw}
H.~Itoyama, S.~Nakajima, {Stability, enhanced gauge symmetry and suppressed
  cosmological constant in 9D heterotic interpolating models}, Nucl. Phys. B
  958 (2020) 115111.
\newblock \href {http://arxiv.org/abs/2003.11217} {\path{arXiv:2003.11217}},
  \href {https://doi.org/10.1016/j.nuclphysb.2020.115111}
  {\path{doi:10.1016/j.nuclphysb.2020.115111}}.

\bibitem{Itoyama:2021fwc}
H.~Itoyama, S.~Nakajima, {Marginal deformations of heterotic interpolating
  models and exponential suppression of the cosmological constant}, Phys. Lett.
  B 816 (2021) 136195.
\newblock \href {http://arxiv.org/abs/2101.10619} {\path{arXiv:2101.10619}},
  \href {https://doi.org/10.1016/j.physletb.2021.136195}
  {\path{doi:10.1016/j.physletb.2021.136195}}.

\bibitem{Itoyama:2021itj}
H.~Itoyama, Y.~Koga, S.~Nakajima, {Target space duality of non-supersymmetric
  string theory}, Nucl. Phys. B 975 (2022) 115667.
\newblock \href {http://arxiv.org/abs/2110.09762} {\path{arXiv:2110.09762}},
  \href {https://doi.org/10.1016/j.nuclphysb.2022.115667}
  {\path{doi:10.1016/j.nuclphysb.2022.115667}}.

\bibitem{Itoyama:2021kxp}
H.~Itoyama, Y.~Koga, S.~Nakajima, {Gauge Symmetry Enhancement by Wilson Lines
  in Twisted Compactification} (6 2021).
\newblock \href {http://arxiv.org/abs/2106.10629} {\path{arXiv:2106.10629}}.

\bibitem{Koga:2022qch}
Y.~Koga, {Interpolation and Exponentially Suppressed Cosmological Constant in
  Non-Supersymmetric Heterotic Strings with General $\mathbb{Z}_{2}$ Twists}
  (12 2022).
\newblock \href {http://arxiv.org/abs/2212.14572} {\path{arXiv:2212.14572}}.

\bibitem{Florakis:2016ani}
I.~Florakis, J.~Rizos, {Chiral Heterotic Strings with Positive Cosmological
  Constant}, Nucl. Phys. B 913 (2016) 495--533.
\newblock \href {http://arxiv.org/abs/1608.04582} {\path{arXiv:1608.04582}},
  \href {https://doi.org/10.1016/j.nuclphysb.2016.09.018}
  {\path{doi:10.1016/j.nuclphysb.2016.09.018}}.

\bibitem{Abel:2017rch}
S.~Abel, R.~J. Stewart, {Exponential suppression of the cosmological constant
  in nonsupersymmetric string vacua at two loops and beyond}, Phys. Rev. D
  96~(10) (2017) 106013.
\newblock \href {http://arxiv.org/abs/1701.06629} {\path{arXiv:1701.06629}},
  \href {https://doi.org/10.1103/PhysRevD.96.106013}
  {\path{doi:10.1103/PhysRevD.96.106013}}.

\bibitem{Abel:2020ldo}
S.~Abel, T.~Coudarchet, H.~Partouche, {On the stability of open-string orbifold
  models with broken supersymmetry}, Nucl. Phys. B 957 (2020) 115100.
\newblock \href {http://arxiv.org/abs/2003.02545} {\path{arXiv:2003.02545}},
  \href {https://doi.org/10.1016/j.nuclphysb.2020.115100}
  {\path{doi:10.1016/j.nuclphysb.2020.115100}}.

\bibitem{Coudarchet:2020sjw}
T.~Coudarchet, H.~Partouche, {One-loop masses of Neumann-Dirichlet open strings
  and boundary-changing vertex operators}, JHEP 12 (2021) 022.
\newblock \href {http://arxiv.org/abs/2011.13725} {\path{arXiv:2011.13725}},
  \href {https://doi.org/10.1007/JHEP12(2021)022}
  {\path{doi:10.1007/JHEP12(2021)022}}.

\bibitem{Kachru:1998hd}
S.~Kachru, J.~Kumar, E.~Silverstein, {Vacuum energy cancellation in a
  nonsupersymmetric string}, Phys. Rev. D 59 (1999) 106004.
\newblock \href {http://arxiv.org/abs/hep-th/9807076}
  {\path{arXiv:hep-th/9807076}}, \href
  {https://doi.org/10.1103/PhysRevD.59.106004}
  {\path{doi:10.1103/PhysRevD.59.106004}}.

\bibitem{Shiu:1998he}
G.~Shiu, S.~H.~H. Tye, {Bose-Fermi degeneracy and duality in nonsupersymmetric
  strings}, Nucl. Phys. B 542 (1999) 45--72.
\newblock \href {http://arxiv.org/abs/hep-th/9808095}
  {\path{arXiv:hep-th/9808095}}, \href
  {https://doi.org/10.1016/S0550-3213(98)00775-5}
  {\path{doi:10.1016/S0550-3213(98)00775-5}}.

\bibitem{Satoh:2015nlc}
Y.~Satoh, Y.~Sugawara, T.~Wada, {Non-supersymmetric Asymmetric Orbifolds with
  Vanishing Cosmological Constant}, JHEP 02 (2016) 184.
\newblock \href {http://arxiv.org/abs/1512.05155} {\path{arXiv:1512.05155}},
  \href {https://doi.org/10.1007/JHEP02(2016)184}
  {\path{doi:10.1007/JHEP02(2016)184}}.

\bibitem{Sugawara:2016lpa}
Y.~Sugawara, T.~Wada, {More on Non-supersymmetric Asymmetric Orbifolds with
  Vanishing Cosmological Constant}, JHEP 08 (2016) 028.
\newblock \href {http://arxiv.org/abs/1605.07021} {\path{arXiv:1605.07021}},
  \href {https://doi.org/10.1007/JHEP08(2016)028}
  {\path{doi:10.1007/JHEP08(2016)028}}.

\bibitem{Dixon:1986jc}
L.~J. Dixon, J.~A. Harvey, C.~Vafa, E.~Witten, {Strings on Orbifolds. 2.},
  Nucl. Phys. B 274 (1986) 285--314.
\newblock \href {https://doi.org/10.1016/0550-3213(86)90287-7}
  {\path{doi:10.1016/0550-3213(86)90287-7}}.

\bibitem{Kawai:1986vd}
H.~Kawai, D.~C. Lewellen, S.~H.~H. Tye, {Classification of Closed Fermionic
  String Models}, Phys. Rev. D 34 (1986) 3794.
\newblock \href {https://doi.org/10.1103/PhysRevD.34.3794}
  {\path{doi:10.1103/PhysRevD.34.3794}}.

\bibitem{Chaudhuri:1995fk}
S.~Chaudhuri, G.~Hockney, J.~D. Lykken, {Maximally supersymmetric string
  theories in D \ensuremath{<} 10}, Phys. Rev. Lett. 75 (1995) 2264--2267.
\newblock \href {http://arxiv.org/abs/hep-th/9505054}
  {\path{arXiv:hep-th/9505054}}, \href
  {https://doi.org/10.1103/PhysRevLett.75.2264}
  {\path{doi:10.1103/PhysRevLett.75.2264}}.

\bibitem{Kawai:1986ah}
H.~Kawai, D.~C. Lewellen, S.~H.~H. Tye, {Construction of Fermionic String
  Models in Four-Dimensions}, Nucl. Phys. B 288 (1987) 1.
\newblock \href {https://doi.org/10.1016/0550-3213(87)90208-2}
  {\path{doi:10.1016/0550-3213(87)90208-2}}.

\bibitem{Font:2021uyw}
A.~Font, B.~Fraiman, M.~Gra\~na, C.~A. N\'u\~nez, H.~Parra De~Freitas,
  {Exploring the landscape of CHL strings on T$^{d}$}, JHEP 08 (2021) 095.
\newblock \href {http://arxiv.org/abs/2104.07131} {\path{arXiv:2104.07131}},
  \href {https://doi.org/10.1007/JHEP08(2021)095}
  {\path{doi:10.1007/JHEP08(2021)095}}.

\bibitem{Partouche:2020swy}
H.~Partouche, B.~De~Vaulchier, {Heterotic orbifolds, reduced rank and $SO(2n +
  1)$ characters}, Int. J. Mod. Phys. A 35~(24) (2020) 2050132.
\newblock \href {http://arxiv.org/abs/2006.08194} {\path{arXiv:2006.08194}},
  \href {https://doi.org/10.1142/S0217751X20501328}
  {\path{doi:10.1142/S0217751X20501328}}.

\bibitem{Narain:1985jj}
K.~S. Narain, {New Heterotic String Theories in Uncompactified Dimensions
  \ensuremath{<} 10}, Phys. Lett. B 169 (1986) 41--46.
\newblock \href {https://doi.org/10.1016/0370-2693(86)90682-9}
  {\path{doi:10.1016/0370-2693(86)90682-9}}.

\bibitem{Narain:1986am}
K.~S. Narain, M.~H. Sarmadi, E.~Witten, {A Note on Toroidal Compactification of
  Heterotic String Theory}, Nucl. Phys. B 279 (1987) 369--379.
\newblock \href {https://doi.org/10.1016/0550-3213(87)90001-0}
  {\path{doi:10.1016/0550-3213(87)90001-0}}.

\bibitem{Dixon:1985jw}
L.~J. Dixon, J.~A. Harvey, C.~Vafa, E.~Witten, {Strings on Orbifolds}, Nucl.
  Phys. B 261 (1985) 678--686.
\newblock \href {https://doi.org/10.1016/0550-3213(85)90593-0}
  {\path{doi:10.1016/0550-3213(85)90593-0}}.

\bibitem{Mikhailov:1998si}
A.~Mikhailov, {Momentum lattice for CHL string}, Nucl. Phys. B 534 (1998)
  612--652.
\newblock \href {http://arxiv.org/abs/hep-th/9806030}
  {\path{arXiv:hep-th/9806030}}, \href
  {https://doi.org/10.1016/S0550-3213(98)00605-1}
  {\path{doi:10.1016/S0550-3213(98)00605-1}}.

\bibitem{Font:2020rsk}
A.~Font, B.~Fraiman, M.~Gra\~na, C.~A. N\'u\~nez, H.~P. De~Freitas, {Exploring
  the landscape of heterotic strings on $T^d$}, JHEP 10 (2020) 194.
\newblock \href {http://arxiv.org/abs/2007.10358} {\path{arXiv:2007.10358}},
  \href {https://doi.org/10.1007/JHEP10(2020)194}
  {\path{doi:10.1007/JHEP10(2020)194}}.

\bibitem{Basile:2018irz}
I.~Basile, J.~Mourad, A.~Sagnotti, {On Classical Stability with Broken
  Supersymmetry}, JHEP 01 (2019) 174.
\newblock \href {http://arxiv.org/abs/1811.11448} {\path{arXiv:1811.11448}},
  \href {https://doi.org/10.1007/JHEP01(2019)174}
  {\path{doi:10.1007/JHEP01(2019)174}}.

\bibitem{Antonelli:2019nar}
R.~Antonelli, I.~Basile, {Brane annihilation in non-supersymmetric strings},
  JHEP 11 (2019) 021.
\newblock \href {http://arxiv.org/abs/1908.04352} {\path{arXiv:1908.04352}},
  \href {https://doi.org/10.1007/JHEP11(2019)021}
  {\path{doi:10.1007/JHEP11(2019)021}}.

\bibitem{Basile:2021vxh}
I.~Basile, {Supersymmetry breaking and stability in string vacua: Brane
  dynamics, bubbles and the swampland}, Riv. Nuovo Cim. 44~(10) (2021)
  499--596.
\newblock \href {http://arxiv.org/abs/2107.02814} {\path{arXiv:2107.02814}},
  \href {https://doi.org/10.1007/s40766-021-00024-9}
  {\path{doi:10.1007/s40766-021-00024-9}}.

\end{thebibliography}

\end{document}